\documentclass{article}
\pdfoutput=1
\usepackage{graphicx} 
\usepackage{subcaption}
\usepackage{amsmath}
\usepackage[style=numeric-comp,sorting=none]{biblatex}
\usepackage{authblk}
\usepackage{cleveref}
\usepackage{mathrsfs}

\DeclareUnicodeCharacter{03C9}{$\omega$}
\DeclareUnicodeCharacter{03B4}{$\delta$}
\DeclareUnicodeCharacter{0394}{$\Delta$}

\title{Phase Stability Analysis of Volume-preserving Algorithms for  Accurate  Single Particle Orbit Simulations in Tokamak Plasmas}

\author[1,2]{Jian Wang\thanks{Corresponding Author. Email address: wang.jian@ipp.ac.cn}}
\author[1]{Xiaodong Zhang}
\author[1]{Lei Ye}
\author[1]{Xingyuan Xu}

\affil[1]{Institute of Plasma Physics, Hefei Institutes of Physical Science, Chinese Academy of Science, Hefei 230031, China}

\affil[2]{University of Science and Technology of China, Hefei 230026, China}

\date{}

\begin{document}

\maketitle

\begin{abstract}
     Second-order Volume-preserving algorithms (VPAs) for simulating charged particle motion in electromagnetic fields have been generalized to a rotating angle formulation by using the matrix decomposition methods. Based on this method, the phase stability of this class of VPAs has been analyzed by using the Discrete Fourier Transformations (DFT) technique. It is found that two prominent VPAs, namely the $G_h^2$ and the Boris algorithm, exhibit optimal phase precision for high-frequency (gyro motion) and low-frequency dynamics (transit/bounce motion), respectively.  These findings have been empirically verified through numerical experiments.  The insights gained from this study enable the selection of an appropriate VPA for practical simulations based on the characteristic frequencies of specific physics problems, which can substantially enhance numerical accuracy and improve computational efficiency for long-term simulations. 
\end{abstract}

\section{Introduction}
    ~~~~~Numerical simulation of the trajectory of a charged particle in a tokamak is a fundamental problem with significant importance within the field.  On one hand, understanding the behavior of individual particles in specific electromagnetic fields can elucidate numerous key physics phenomena.  On the other hand, the aggregation of single-particle orbits lays the groundwork for first-principle, self-consistent large-scale numerical simulations, such as in various particle-in-cell (PIC) \cite{CHEN2003-PIC,LANTI2020-Orb5} and semi-Lagrangian (SL) \cite{GYSELA} codes.

    The motion of a charged particle in tokamak plasma consist of a fast gyro-motion and a slow drift-motion of the gyro-center. For low frequency problems with $\omega\ll\omega_{c}$ , such as drift-waves and shear Alfvén waves, it is sufficient to employ the gyrokinetic model \cite{Brizard2007-GYRO} and trace the gyro-center trajectory by averaging out the gyro-angle, reducing the particle dynamics from 6D to 5D.  Here, $\omega$ and $\omega_c$  represent the characteristic frequency of waves and gyro-motion, respectively. However, for the problems with $\omega\ge\omega_{c}$, such as radio frequency (RF) waves \cite{KuleyPoP2013-RF,KuleyPop2015-RF}, high-frequency turbulence \cite{raeth2023highfrequencynongyrokineticturbulence} and ion cyclotron emission (ICE) \cite{PoP2022-ICE}, the full kinetic model must be utilized, which requires the computation of the full orbit of particles. Moreover,  in the gyrokinetic simulations of edge/pedestal plasma, there has also been a paradigm shift towards utilizing full kinetic ions in place of gyrokinetic ions \cite{ChenPoP2009,LinPPCF2011-GeFi}. This approach significantly extends the applicability of  the numerical models across a broader spectrum of time-space scales. 
       
    The Boris algorithm \cite{Boris,Computer_Simulations_Using_Particles,Plasma_Physics_via_Computer_Simulation} has become the de facto standard of the explicit integration schemes for calculating the full particle orbits in tokamak plasma. Despite its relatively lower precision of truncation error (2nd-order accuracy, in comparison to the 4th-order Runge-Kutta scheme) in a single time step, the Boris algorithm has inherent  robust conservation properties across long temporal scales \cite{ParkerJCP1991-Boris,Stoltz2002-Boris,Penn2003-Boris}, which are crucial for the numerical investigation of  the multi-time-scale nature of plasma physics in tokamak.  It has been pointed out that  a key factor in its success is the algorithm's ability to conserve phase space volume, essential for all Hamiltonian systems \cite{QinPoP2013}.  Such numerical schemes, known as Volume-Preserving Algorithms (VPAs), can be systematically derived through Lie algebraic methods \cite{ZhangCCP2015-LieAlgebra}. Another volume-preserving algorithm, which is firstly introduced in \cite{HeJCP2016-VPA} and herein referred to as $G_h^2$ , has also been developed by modifying the magnetic-field-induced rotation angle of particle velocity in the Boris algorithm. While $G_h^2$  exhibits a minor disadvantage in the accuracy of adiabatic invariant, it possesses a faster convergence rate for numerical solutions compared to the Boris algorithm in  a static electromagnetic field. Besides, advanced volume-preserving methods of higher precision have been developed and applied in both non-relativistic and relativistic dynamics of charged particles \cite{HeJCP2016-Higher_order_VPA,HePoP2016-Relativistic_VPA,WangPoP2016-Runaway_Electrons}. These methods have delivered highly accurate results in long-term simulations.

    Although the theoretical framework of VPAs has reached a sophisticated level, the main focus has primarily been on the conservative properties of VPAs in long-time simulations. It is known that the wave-particle resonances between charged particles and a spectrum of electromagnetic waves across different frequencies play an essential role in most physics phenomena in tokamak plasma, such as RF heating and Energetic particle (EP)-driven instabilities. Therefore, numerical precision in the particle phase in given fields is of great importance for numerical investigations of these issues. However, the phase stability of various VPAs has not been extensively explored.  

    In this work, VPAs for simulating charged particle motion in electromagnetic fields have been generalized to a rotating angle formulation using matrix decomposition methods. Based on this method, the phase stability of a class of VPAs has been analyzed using Discrete Fourier Transformations (DFT) technique. The theoretical phase stability analysis presented here proves that the Boris algorithm and  $G_h^2$  are optimally suited for computing low and high-frequency components, respectively, within this class of second-order VPAs. Numerical simulations of charged particle trajectories within a typical tokamak toroidal magnetic field have been executed to verify the analytical predictions, demonstrating different performances in calculating various scales of motion by the two representative VPAs. These new findings can enable the selection of more appropriate numerical integral schemes in constructing full particle orbits, potentially with larger time step sizes, based on the characteristic frequency of the physical problem at hand.  Such selection can also substantially reduce computational time and improve the efficiency of long-term simulations.

This paper is organized as follows. Section II delineates the generalization of a series of VPAs in matrix notation. Sec.III presents the theoretical phase stability analysis of the volume-preserving algorithms across various frequencies. Compared in Sec.IV are the numerical results of the charged particle trajectories in a typical Tokamak toroidal magnetic field by the Boris algorithm and $G_h^2$ respectively. Finally, Sec.V concludes the paper.

\section{Generalization of Volume-Preserving Algorithms}
~~~~This section is dedicated to the generalization of VPAs, adhering to the definitions established in \cite{QinPoP2013}. The formulation of these algorithms is articulated through matrix notation, which will facilitate the subsequent phase stability analysis.

The motion of charged particles in an electromagnetic field $\vec E=(E^x,E^y,E^z)^T$ and $\vec B=(B^x,B^y,B^z)^T$ is governed by the Lorentz-Newton equation

$$m \frac{d\vec v}{dt}=q(\vec v \times \vec B+\vec E) \eqno{(1.a)} $$
$$\frac{d\vec r}{dt}=\vec v \eqno{(1.b)}$$
with $m$ the mass, $q$ the electric charge, and $\vec r=(x,y,z)^T$, $\vec v=(v^x,v^y,v^z)^T$ the position and velocity of the charged particle under Cartesian coordinates. To facilitate the discussion, we will normalize the magnetic field $\vec B$, electric field $\vec E$, velocity variable $\vec v$, position variable $\vec r$ time variable $t$ by basic quantities 
$$
B_{ref} = B_0,v_{ref} = v_0
\eqno{(2.a)} 
$$
with $B_0$ the magnetic field strength on the magnetic axis, $v_0$ the initial velocity magnitude of the particle. And derived quantities are given by
$$
E_{ref}=B_{ref} v_{ref}=B_0v_0,t_{ref}=\frac{m}{qB_{ref}}=\frac{m}{qB_0},r_{ref}=v_{ref} t_{ref} = \frac{mv_0}{qB_0}
\eqno{(2.b)} 
$$
Replacing $\vec B, \vec E,\vec v,\vec r$ and $t$  by $\frac{\vec B}{B_{ref}},\frac{\vec E}{E_{ref}},\frac{\vec v}{v_{ref}},\frac{\vec r}{r_{ref}}$ and $\frac{t}{t_{ref}}$ in equations (1) yields
$$\frac{d\vec v}{dt}=\vec v \times \vec B+\vec E \eqno{(3.a)} $$
$$\frac{d\vec r}{dt}=\vec v \eqno{(3.b)}$$
Throughout the discourse in Section 2 and Section 3, we shall persistently utilize the above normalized form. The Lorentz force term $\vec v \times \vec B$  can be written in matrix form as

$$\frac{d\vec v}{dt}=\mathscr{B} \vec v+\vec E \eqno{(4.a)} $$
Here, the real skew-symmetric matrix $\mathscr{B}$ is given by 
$$
\mathscr{B}=
\begin{bmatrix}
 0 & B^z & -B^y\\
 -B^z & 0 & B^x\\
 B^y & -B^x & 0\\
\end{bmatrix}
\eqno{(4.b)}
$$
Since $\mathscr{B}$ is a real skew-symmetric matrix, it can be diagonalized by a unitary matrix $P$
$$
\mathscr{B}=P\Lambda P^* \eqno{(5.a)}
$$
Here, the unitary matrix $P$ and the diagonal matrix $\Lambda$ are given by
$$
P=\frac{1}{B}
\begin{bmatrix}
 B^x & \frac{-B^x B^y - B^z B i}{\sqrt{2((B^x)^2+(B^z)^2)}} & \frac{B^x B^y - B^z B i}{\sqrt{2((B^x)^2+(B^z)^2)}}\\
 B^y & \frac{\sqrt{(B^x)^2+(B^z)^2}}{\sqrt{2}} & -\frac{\sqrt{(B^x)^2+(B^z)^2}}{\sqrt{2}}\\
 B^z & \frac{-B^y B^z + B^x B i}{\sqrt{2((B^x)^2+(B^z)^2)}} & \frac{B^y B^z + B^x B i}{\sqrt{2((B^x)^2+(B^z)^2)}}\\
\end{bmatrix}
\eqno{(5.b)}
$$
$$
\Lambda=\text{diag}(0,Bi,-Bi)\eqno{(5.c)}
$$
with $P^*$ the conjugate transpose matrix of $P$, $B=\sqrt{(B^x)^2+(B^y)^2+(B^z)^2}$ the magnetic field strength. 

Now we consider the numerical integral schemes of (3). Let $\Delta t$ denotes the fixed time step size , the subscript $k$  represents variables in the $k-th$  time step and $z_k=(\vec r_k, \vec v_k)=(\vec r ((k+\frac{1}{2})\Delta t),\vec v (k\Delta t))$ denotes coordinates of the  time step in phase space. The recurrence relation from $z_k$ to $z_{k+1}$ in numerical algorithms generates a one-step map $\psi$ 
$$
\psi:z_k=(\vec r_k, \vec v_k) \rightarrow z_{k+1}=(\vec r_{k+1}, \vec v_{k+1})
\eqno{(6)}
$$
Algorithms satisfying $|\frac{\partial \psi}{\partial z_k}|=1$  for arbitrary $k$ conserve the phase space volume in each time step, thus are referred to as volume-preserving algorithms \cite{QinPoP2013}. The well-known Boris method, as a typical example of VPA, handles the electric and magnetic forces separately
$$
\vec v^- = \vec v_k +\frac{1}{2} \Delta t \cdot \vec E_k 
\eqno{(7.a)}
$$
$$
\vec v^+ =\vec v^- + \frac{1}{2}  \Delta t \cdot (\vec v^+ +\vec v^-) \times \vec B_k
\eqno{(7.b)}
$$
$$
\vec v_{k+1}=\vec v^+ + \frac{1}{2}  \Delta t \cdot \vec E_k 
\eqno{(7.c)}
$$
$$
\vec r_{k+1}=\vec r_k+\Delta t \cdot \vec v_{k+1}
\eqno{(7.d)}
$$
with $\vec E_k=\vec E((k + \frac{1}{2})\cdot \Delta t),\vec B_k=\vec B((k + \frac{1}{2})\cdot \Delta t)$ the electromagnetic field of the $k-th$ time step.  Equations (7) can be similarly written in matrix form as
$$
\vec v_{k+1}=R_k^B \vec v_k + \frac{\Delta t}{2}(I+R_k^B)\vec E_k
\eqno{(8.a)}
$$
$$
\vec r_{k+1}=\vec r_k +{\Delta t}\cdot R_k^B \vec v_k+\frac{\Delta t^2}{2}(I+R_k^B)\vec E_k
\eqno{(8.b)}
$$
Here, the rotation matrix of the Boris algorithm in the $k-th$  time step $R_k^B$ is given by
$$
R_k^B=(I-\frac{\Delta t}{2} {\mathscr{B}_k})^{-1} (I+\frac{\Delta t}{2} {\mathscr{B}_k})
\eqno{(8.c)}
$$
Substituting equations (5) into (8.c) yields
$$
R_k^B=P_k \Lambda_k^B P_k^*
\eqno{(9.a)}
$$
Here, the diagonal matrix $\Lambda_k^B$ are given by
$$
\Lambda_k^B=\text{diag}(1,\text{exp}(\theta_k^B \cdot i),\text{exp}(-\theta_k^B \cdot i))
\eqno{(9.b)}
$$
and $\theta_k^B=2\text{arctan}(\frac{1}{2} {B_k} \cdot \Delta t)$ is the magnetic-field-induced rotation angle of the velocity variable $\vec v_k$ in the k-th time step, which satisfies the following condition of consistency
$$
\lim_{\Delta t\to 0} \frac{\theta_k^B}{{B_k}\cdot \Delta t}=1 
\eqno{(9.c)}
$$

According to Ref. \cite{QinPoP2013}, the volume-preserving condition is identical to $\text{det}(R_k^B)=1$ for arbitraty $k$ . Ref. \cite{QinPoP2013} provides proves by the theory of Cayley transformations, while a different perspective is presented in this paper
$$
\text{det}(R_k^B)=|\text{det}(P_k)\text{det}(\Lambda_k^B)\text{det}(P_k^*)|=|\text{det}(P_k)||\text{det}(\Lambda_k^B)||\text{det}(P_k^*)|
\eqno{(10)}
$$
Since $P_k$ is a unitary matrix, $|\text{det}(P_k)|=|\text{det}(P_k^*)|=1$. From equation (9.b), we have
$$
|\text{det}(\Lambda_k^B)|=1\cdot |\text{exp}(\theta_k^B \cdot i)|\cdot|\text{exp}(-\theta_k^B \cdot i)|=1
\eqno{(11)}
$$
Thus $\text{det}(R_k^B)=1$ is easily obtained. 
Noted that the volume-preserving property remains valid for any other rotation angle $\theta_k$ satisfying the condition of consistency. A class of volume-preserving algorithms can therefore be easily generalized by a similar methodology.

\textbf{Generalization of a class of volume-preserving algorithms.} A class of volume-preserving algorithms can be derived from the following format
$$
\vec v_{k+1}=R_k \vec v_k + \frac{\Delta t}{2}(I+R_k)\vec E_k
\eqno{(12.a)}
$$
$$
\vec r_{k+1}=\vec r_k +{\Delta t}\cdot R_k \vec v_k+\frac{\Delta t^2}{2}(I+R_k){\vec E_k}
\eqno{(12.b)}
$$
Here,
$$
R_k=P_k \Lambda_k P_k^*
\eqno{(12.c)}
$$
$$
\Lambda_k = \text{diag}(1,\text{exp}(\theta_k \cdot i),\text{exp}(-\theta_k \cdot i))
\eqno{(12.d)}
$$
and $\theta_k$ is the magnetic-field-induced rotation angle of the velocity variable $\vec v_k$ in the $k-th$ time step, which needs to satisfy the following condition of consistency
$$
\lim_{\Delta  t\to 0} \frac{\theta_k}{{B_k}\cdot \Delta t}=1 
\eqno{(12.e)}
$$

A variety of volume-preserving algorithms can be obtained by considering various values of the magnetic-field-induced rotation angle $\theta$ as a function of the time step size $\Delta t$ satisfying equation (12.e). A straightforward approach is to take $\theta=B\cdot \Delta t$. The corresponding algorithm has already been derived and is referred to as $G_h^2$ in \cite{HeJCP2016-VPA} by Lie algebra and represented in exponential matrix form. We will consistently use $G_h^2$ to represent this algorithm in the subsequent text.

\section{Phase Stability Analysis of VPAs in Various Frequencies}

~~~~In this section, we conduct a theoretical analysis of the phase stability of the volume-preserving algorithms constructed by equations (12) across a spectrum of frequencies by the Discrete Fourier Transformations (DFT)
$$
\vec f (\omega)=\frac{1}{N} \sum_{m=0}^{N-1} \vec r (t^m) \text{exp}\left(-\frac{2\pi \omega}{T} t^m \cdot i\right),0 \le \omega \le N-1
\eqno{(13)}
$$
and its inverse transformation is
$$
\vec r (t^n) = \sum_{\omega =0}^{N-1} \vec f (\omega) \text{exp}\left(\frac{2\pi \omega}{T} t^n \cdot i\right),0 \le n \le N-1
\eqno{(14)}
$$
with $\Delta t$ the time step size, $T$ the total simulation time (all variables in this section remain normalized  as stated in Section 2), $N=\frac{T}{\Delta t}$ the total number of time grids and $t^m=(m+\frac{1}{2})\cdot \Delta t$ the time of position variable $\vec r$ in the $m-th$ time step. From equation (14), the position variable $\vec r$ can be interpreted as a linear combination of trigonometric functions, each with a distinct frequency $\frac{2\pi \omega}{T}$ and an associated DFT coefficient $\vec f(\omega)$. Since the slow-scale guiding center motions and the fast-scale cyclotron motions can be regarded as low and high frequency components of $\vec r$, the phase stability of the algorithms can be evaluated through the convergence rate of their respective DFT coefficients $\vec f(\omega)$ as the time step size $\Delta t$ vanishes. 

We make the following assumptions for further derivations

(1). The total simulation time $T$ takes  the period of the slow-scale drift motion, which is much larger than that of the cyclotron period, i.e. $T$ is considerably large and $\frac{1}{T}=O(\epsilon)$ is considered to be an infinitesimal quantity. 

(2). The influence of the electric field  is excluded, namely $\vec E=(0,0,0)^T$. In the absence of the electric field, the energy is strictly preserved by volume-preserving algorithms. From equation (12.a), we have 
$$
||\vec v_{k+1}||_2=||R_k \vec v_{k}||_2=||P_k \Lambda_k P_k^*\vec v_{k}||_2
\eqno{(15)}
$$
Since $P_k$,$\Lambda_k$ and $P_k^*$ are all unitary matrices, their corresponding linear transformations conserve the 2-norm of vectors. Thus we have  $||\vec v_{k}||_2=||\vec v_{k-1}||_2=...=||\vec v_0||_2=v_0$, with $v_0$ the initial magnitude of velocity. 

(3). The direction of the magnetic field remains unaltered, namely $\vec B=B(\vec r)\vec e_0$,which means the corresponding unitary matrix  $P$ is held constant.  The variation of the magnetic field within a single time step $\Delta B$ needs to satisfy the following condition of approximation
$$
\Delta B=O(\epsilon^{1+\alpha}),\alpha>0
\eqno{(16)}
$$
This stipulation elucidates that $\Delta B$ is an infinitesimal of a higher order compared to $\epsilon$.
Under these conditions, the numerical solutions $\vec v^m,\vec r^m$ in the m-th time step are readily obtained
$$
\vec v_m= \prod_{n=1}^{m} R_n \vec v_0 =P \text{diag} \left(1,\prod_{n=1}^{m} \lambda_{2,n},\prod_{n=1}^{m} \lambda_{3,n}\right) P^* \vec v_0
\eqno{(17.a)}
$$
$$
\vec r_m = \vec r_0 + \frac{1}{2}\vec v_0 \cdot \Delta t + \sum_{j=1}^m \vec v_j \cdot \Delta t
\eqno{(17.b)}
$$
Here $\lambda_{2,n}=\text{exp}(\theta_n \cdot i), \lambda_{3,n}=\text{exp}(-\theta_n \cdot i)$. From (13), we have 
$$
\vec f(\omega)= \frac{\Delta t}{T} \sum_{m=0}^{N-1} \vec r_m \text{exp} \left(-k(m+\frac{1}{2})\Delta t\cdot i) \right) \
\eqno{(18)}
$$
with $k=\frac{2\pi \omega}{T}$. Substituting equations (17) into (18) yields
$$
\vec f(\omega)=  \frac{\Delta t}{T} \frac{\text{exp}(-\frac{1}{2}k\Delta t\cdot i)}{1-\text{exp}(-k\Delta t\cdot i)} P \text{diag}(\lambda_1,\lambda_2,\lambda_3) P^* \vec v_0
\eqno{(19.a)}
$$
Given the above condition (16), $\lambda_1,\lambda_2,\lambda_3$ can be approximated as
$$
\lambda_1=-N\Delta t=-T
\eqno{(19.b)}
$$
$$
\lambda_2=\left[ \left(\prod_{m=2}^{N}\lambda_{2,m}-1 \right) \left (\frac{\lambda_{2,1}}{1-\lambda_{2,1}}-\frac{\lambda_{2,1}\text{exp}(-k\Delta t \cdot i)}{1-\lambda_{2,1}\text{exp}(-k\Delta t \cdot i)}\right )+O(\epsilon^{\alpha}) \right]\Delta t
\eqno{(19.c)}
$$
$$
\lambda_3=\left[\left(\prod_{m=2}^{N}\lambda_{3,m}-1\right)\left(\frac{\lambda_{3,1}}{1-\lambda_{3,1}}-\frac{\lambda_{3,1}\text{exp}(-k\Delta t \cdot i)}{1-\lambda_{3,1}\text{exp}(-k\Delta t \cdot i)}\right)+O(\epsilon^{\alpha})\right]\Delta t
\eqno{(19.d)}
$$
with $\lambda_{2,1}$ the initial value of $\lambda_{2,n}$, $\lambda_{3,1}$ the initial value of $\lambda_{3,n}$. See specific derivations of approximation in appendix. $\lambda_2,\lambda_3$ are associated with the magnetic-field-induced rotation angle $\theta$, while the remaining variables in equations (19) are irrelevant and will therefore be disregarded in the subsequent discussions. We will exclusively concentrate on the convergence rate of $\lambda_2$ in various volume-preserving algorithms and their corresponding rotation angle $\theta$. $\lambda_2$ can be reduced to the following form
$$
\lambda_2=\left(\prod_{m=2}^{N}\lambda_{2,m}-1\right)[h(k,\Delta t)+O(\epsilon^{\alpha}\Delta t)]\cdot i
\eqno{(20.a)}
$$
Here, $h(k,\Delta t)$ is given by
$$
h(k,\Delta t)=\left(\frac{2\text{tan}\frac{\theta_1}{2}}{\Delta t}\right) ^{-1}-\left(\frac{2\text{tan}\frac{\theta_1-k\Delta t}{2}}{\Delta t}\right) ^{-1}
\eqno{(20.b)}
$$
Since $(\prod_{m=2}^{N}\lambda_{2,m}-1)\sim O(1)$, it does not substantially influence the convergence rate of $\lambda_2$. Conversely, $h(k,\Delta t)$ emerges as the primary focus of our attention.  Furthermore, given the periodic nature of the problem, we shall omit the time-dependent subscripts in $h(k,\Delta t)$.   
For the case of low frequency components associated with slow-scale guiding center motions, i.e. $\omega \sim O(1)$, we have $k=\frac{2\pi \omega}{T}\sim O(\frac{1}{T}) \sim O(\epsilon)$  and $\lim_{\Delta t->0}h(k,\Delta t)=0$ . Equation (20.b) is therefore reduced to
$$
h(k,\Delta t)=\left(\frac{2\text{tan}\frac{\theta}{2}}{\Delta t}\right) ^{-1}-\left(\frac{2\text{tan}\frac{\theta}{2}}{\Delta t}+O(\epsilon)\right) ^{-1}
\eqno{(21)}
$$
The rotation angle of the Boris algorithm $\theta_B=2\text{arctan}(\frac{1}{2} B \Delta t)$ yields
$$
h_B(k,\Delta t)=\frac{1}{B^2} O(\epsilon)
\eqno{(22)}
$$
Any other volume-preserving algorithm and its corresponding rotation angle satisfying equation (12.e), i.e. $\theta=2\text{arctan}(\frac{1}{2} B \Delta t)+c(\Delta t^2)+O(\Delta t^3) $ with $c$ an arbitrary constant, yields the following result by calculating the Taylor expansion of $2\text{tan}(\frac{\theta}{2})$
$$
h(k,\Delta t)=\frac{1-2\frac{c}{B}\Delta t+O(\Delta t^2)}{B^2} O(\epsilon)=\frac{1+O(\Delta t)}{B^2} O(\epsilon)
\eqno{(23)}
$$
Equations (22) and (23) demonstrate that the Boris algorithm exhibits the fastest convergence rate of $\lambda_2$ than other volume-preserving methods, due to the fact that it eliminates the first-order and higher-order terms of $\Delta t$ in $h(k,\Delta t)$. Consequently, the Boris algorithm stands as the most effective scheme for calculating slow-scale guiding center motions within this series of volume-preserving algorithms.

Now we consider the case of large $\omega$ related to fast scale cyclotron motions, i.e. $\omega=B \cdot \frac{T}{2\pi}$ and $k=\frac{2\pi \omega}{T}=B$ and $\lim_{\Delta t->0}\frac{1}{h(k,\Delta t)}=0$. Similarly, the rotation angle of $G_h^2$, i.e. $\theta_{G_h^2}=B\Delta t$ yields
$$
\frac{1}{h_{G_h^2}(k,\Delta t)}=0
\eqno{(24)}
$$
Any other volume-preserving algorithm and its corresponding rotation angle satisfying equation (12.e), i.e. $\theta=B\Delta t +c(\Delta t^2)+O(\Delta t^3) $ with $c$ an arbitrary constant, yields the following result by similarly calculating the Taylor expansion of $2\text{tan}\frac{\theta-k\Delta t}{2}$
$$
\frac{1}{h(k,\Delta t)}=-c\Delta t+O(\Delta t^2)=O(\Delta t)
\eqno{(25)}
$$
Therefore, $G_h^2$ proves to be more efficient than other methods in computing fast-scale cyclotron motions in long-term calculations. 

The derivation process outlined above can be succinctly interpreted to mean that the Boris algorithm and $G_h^2$ converge towards the singularities of $h(k,\Delta t)$ at distinct values of $k$, corresponding respectively to the first term $(\frac{2\text{tan}\frac{\theta}{2}}{\Delta t})^{-1}$ and the second term $(\frac{2\text{tan}\frac{\theta-k\Delta t}{2}}{\Delta t})^{-1}$ of $h(k,\Delta t)$. The same analytical procedure can be replicated for $\lambda_3$. 

Until now, our theoretical investigations have successfully identified the most efficient numerical schemes for generating slow-scale guiding center motions (the Boris algorithm) and fast-scale cyclotron motions ($G_h^2$) among the spectrum of volume-preserving algorithms delineated by equations (12). Numerical experiments and comparisons between the Boris algorithm and $G_h^2$ will be conducted in the next section. The explicit expression of their corresponding rotation matrix $R_B$ and $R_{G_h^2}$ are given by
$$
R_B=\frac{1}{4+\theta_x^2}
\begin{bmatrix}
 4+\theta_x^2-\theta_y^2-\theta_z^2 & 2\theta_x\theta_y+4\theta_z & 2\theta_x\theta_z-4\theta_y \\
 2\theta_x\theta_y-4\theta_z &  4+\theta_y^2-\theta_x^2-\theta_z^2 & 2\theta_y\theta_z+4\theta_x \\
 2\theta_x\theta_z+4\theta_y & 2\theta_y\theta_z-4\theta_x & 4+\theta_z^2-\theta_x^2-\theta_y^2\\
\end{bmatrix}
\eqno{(26.a)}
$$
$$
R_{G_h^2}=\frac{1}{\theta^2}
\begin{bmatrix}
 \theta_x^2+(\theta_y^2+\theta_z^2)cos\theta & \theta_x \theta_y (1-cos\theta)+\theta_z \theta sin\theta & \theta_x \theta_z (1-cos\theta)-\theta_y \theta sin\theta \\
 \theta_x \theta_y (1-cos\theta)-\theta_z \theta sin\theta & \theta_y^2+(\theta_x^2+\theta_z^2)cos\theta & \theta_y \theta_z (1-cos\theta)+\theta_x \theta sin\theta \\
 \theta_x \theta_z (1-cos\theta)+\theta_y \theta sin\theta & \theta_y \theta_z (1-cos\theta)-\theta_x \theta sin\theta & \theta_z^2+(\theta_x^2+\theta_y^2)cos\theta\\
\end{bmatrix}
\eqno{(26.b)}
$$
Here
$$
\theta=B\Delta t,\theta_x=B_x\Delta t,\theta_y=B_y\Delta t,\theta_z=B_z\Delta t
\eqno{(26.c)}
$$

\section{Numerical Experiments}
\begin{figure}
\centering
\begin{subfigure}[b]{0.49\textwidth}
  \includegraphics[width=\textwidth]{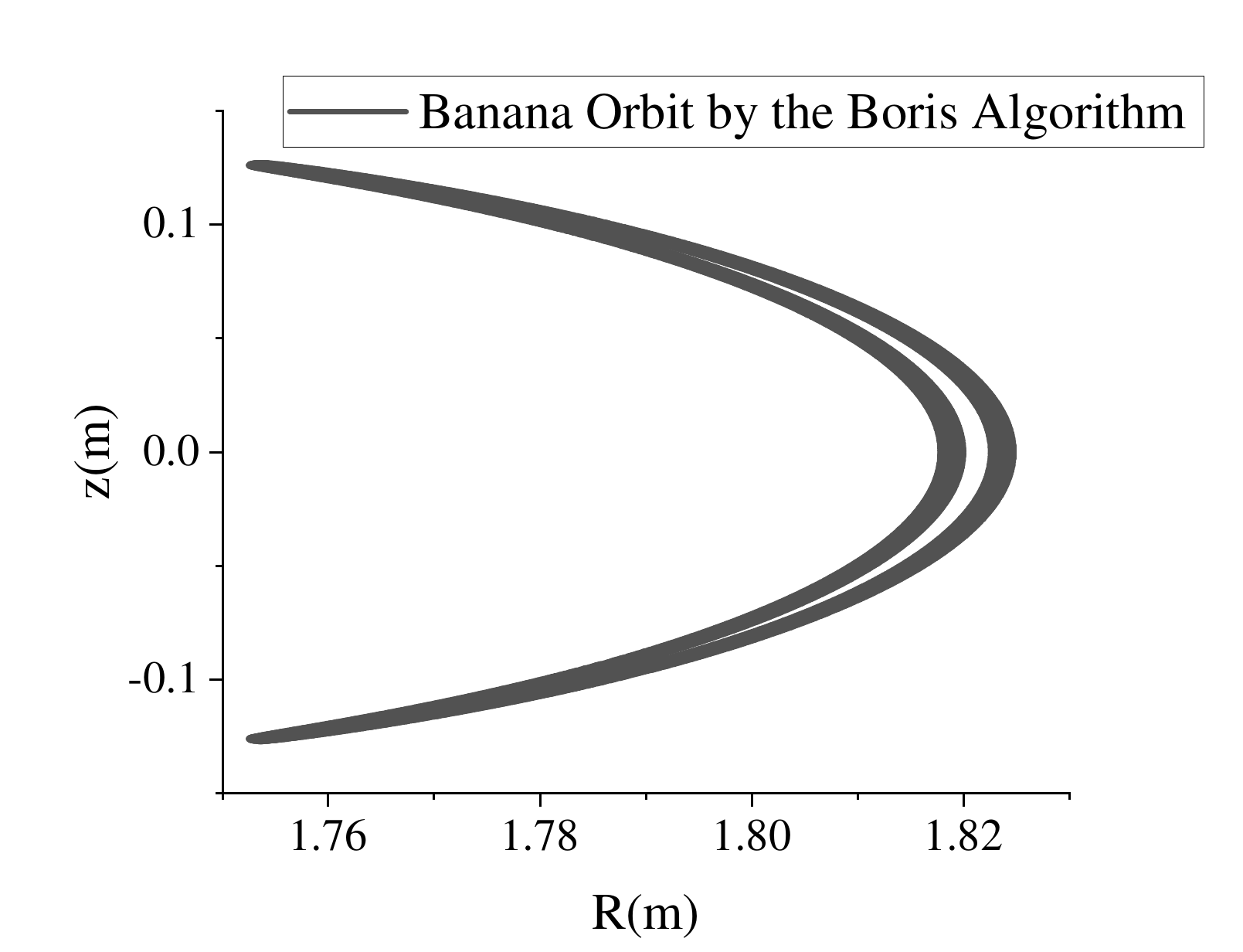}
  \caption{}
  \label{fig 1-a}
\end{subfigure}
\hfill 
\begin{subfigure}[b]{0.49\textwidth}
  \includegraphics[width=\textwidth]{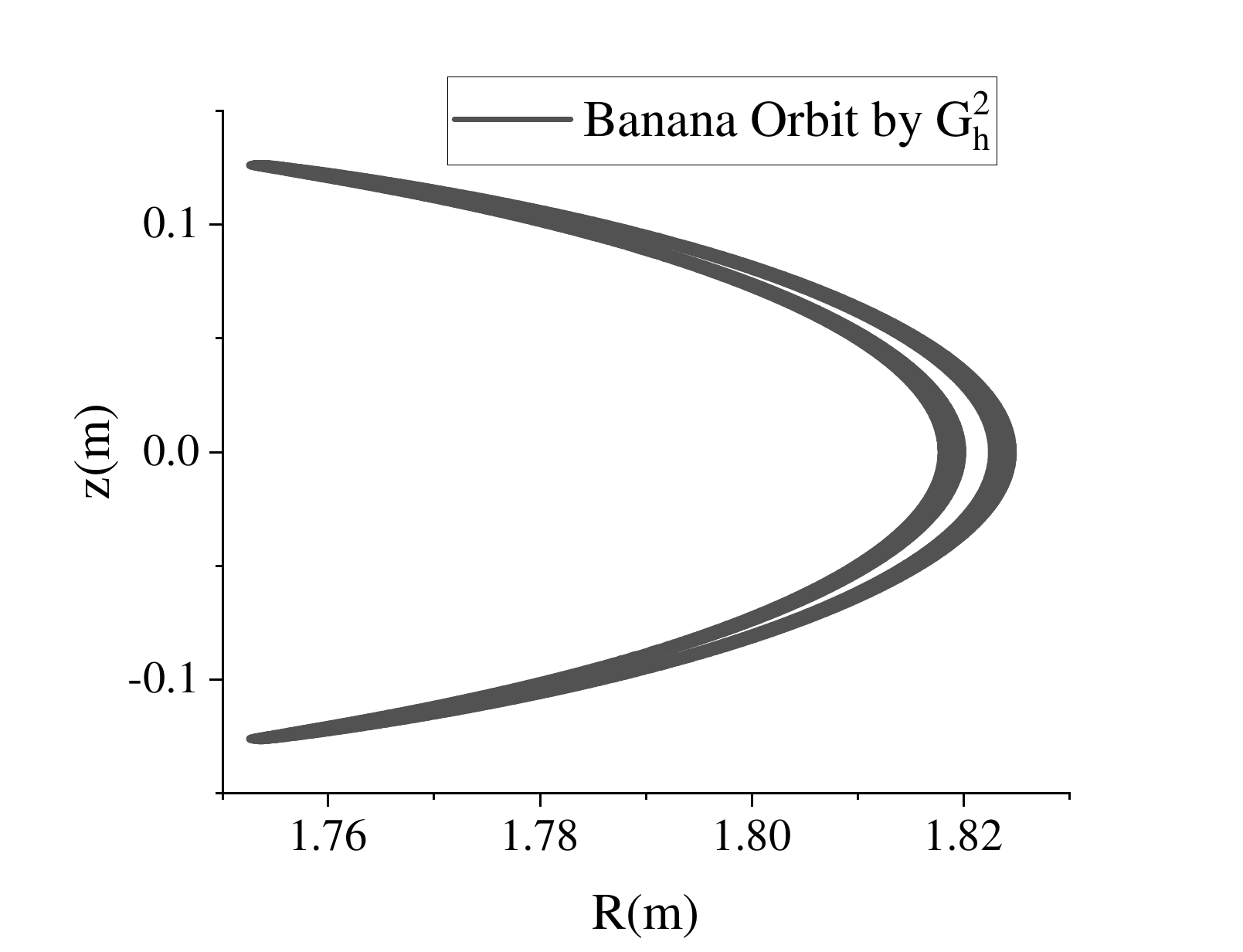}
  \caption{}
  \label{fig 1-b}
\end{subfigure}

\caption{Numerical results with initial conditions of banana orbit. The time step size is $\omega_{c0} \Delta t=0.1$, and the time integration interval is $[0,T_0]$ with $\omega_{c0}T_0=2.54 \times 10^4$ which is approximately one period of the slow-scale motions (banana period). The banana orbit is correctly obtained by both algorithms.}
\label{fig 1}
\end{figure}

\begin{figure}
\centering
\begin{subfigure}[b]{0.49\textwidth}
  \includegraphics[width=\textwidth]{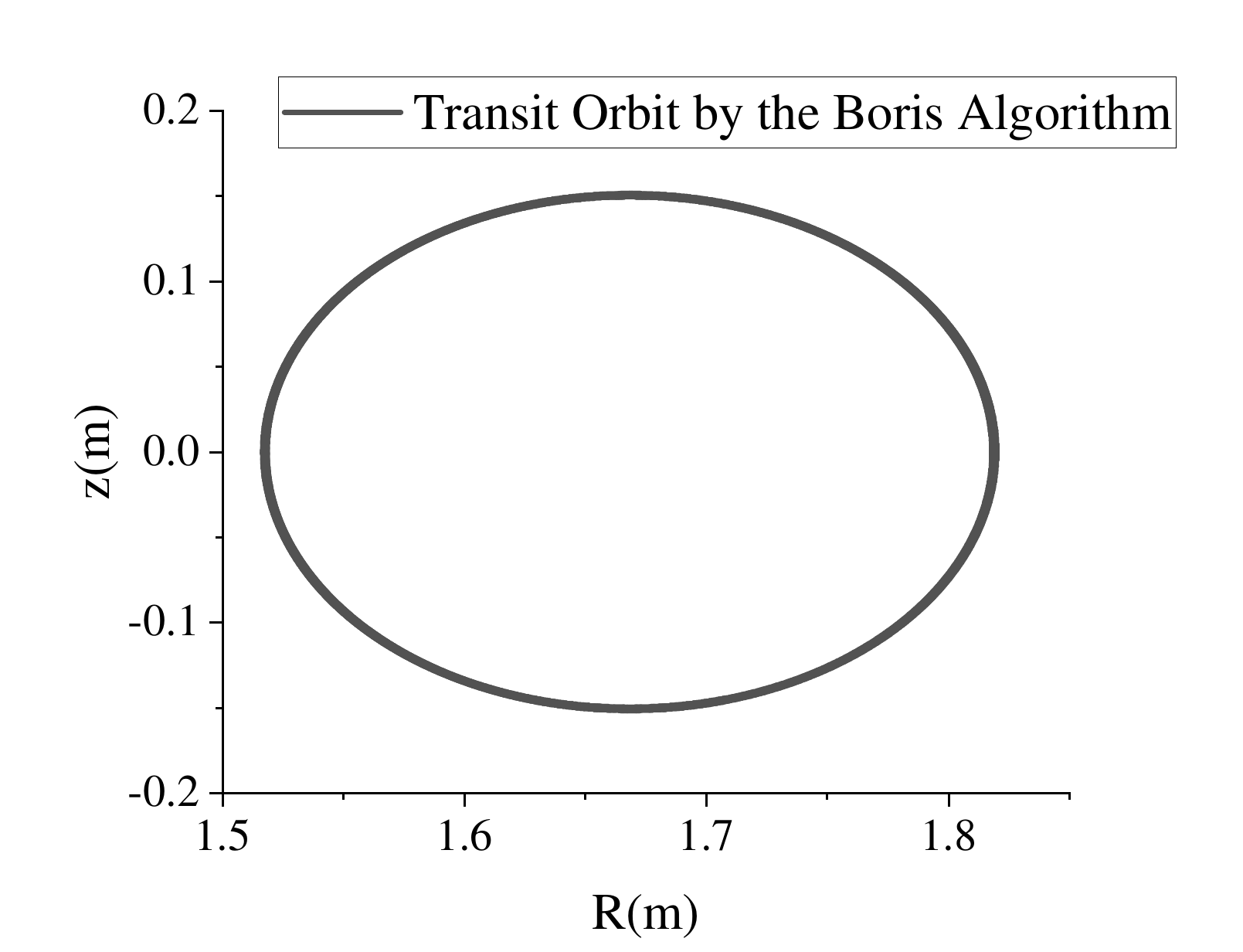}
  \caption{}
  \label{fig 2-a}
\end{subfigure}
\hfill 
\begin{subfigure}[b]{0.49\textwidth}
  \includegraphics[width=\textwidth]{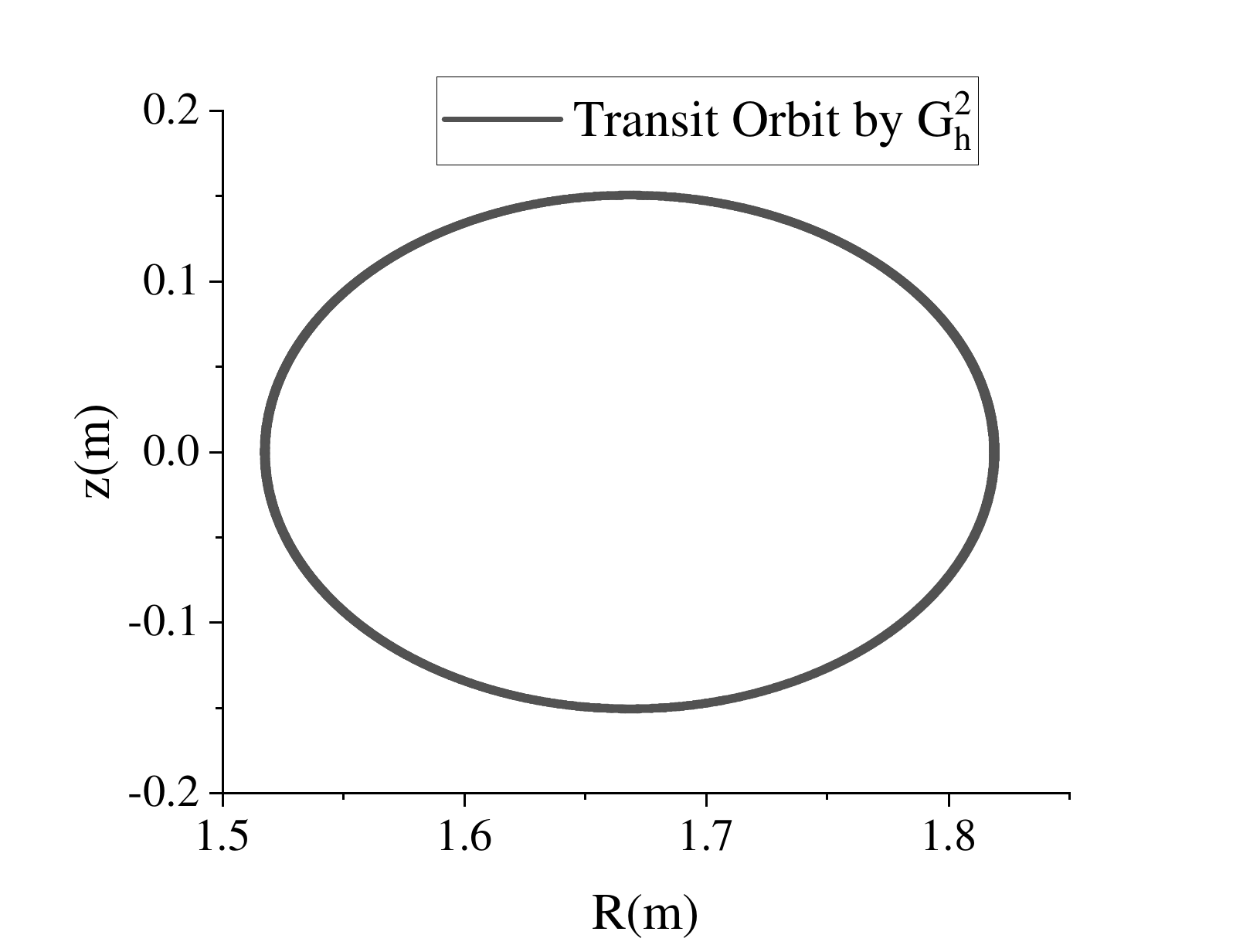}
  \caption{}
  \label{fig 2-b}
\end{subfigure}

\caption{Numerical results with initial conditions of transit orbit. The time step size is $\omega_{c0} \Delta t=0.1$, and the time integration interval is $[0,T_1]$ with $\omega_{c0}T_1=1.38 \times 10^4$ which is approximately one period of the slow-scale motions (transit period). The transit orbit is correctly obtained by both algorithms.}
\label{fig 2}
\end{figure}

~~~~In this section, we numerically test two representative volume-preserving algorithms (the Boris algorithm and $G_h^2$), and verify their properties in calculating fast and slow scale motions. The variables will no longer be normalized in the subsequent text. Instead, the time variables will be qualified by the gyro-frequency of an ion in a certain magnetic field $B_0=1T$
$$
\omega_{c0}=\frac{eB_0}{m_i}=9.57\times 10^7 s^{-1}
\eqno{(27)}
$$
We initially disregard the impact of the electric field, i.e. $\vec E=(0,0,0)^T$. Consider the motion of a single ion in a toroidal magnetic field with magnetic field strength on the magnetic axis $B_{axis}=2T$, major radius $R_0=1.67m$, minor radius $a=0.6m$, and the safety factor 
$$
q=2.52(\frac{r}{a})^2-0.16(\frac{r}{a})+0.86
\eqno{(28)}
$$
with $r=\sqrt{(\sqrt{x^2+y^2}-R_0)^2+z^2}$. These parameters can be referenced in \cite{GorlerPoP2016-GYRO}. 
The magnetic field  in the torodial coordinates $(r,\theta,\phi)$ is expressed as $\vec B=B_{\theta} \vec e_{\theta}+B_{\phi} \vec e_{\phi}$ with
$$
B_{\phi}=\frac{B_{axis}R_0}{R_0+r \text{cos}\theta},B_{\theta}=\frac{rB_{\phi}}{qR_0}
\eqno{(29)}
$$
To apply the volume-preserving algorithms, we transform the toroidal magnetic field $\vec B$ into the Cartesian coordinates $(x,y,z)$ which is
$$
B_x=-B_{\phi} \text{sin}\phi-B_{\theta} \text{sin}\theta \text{cos}\phi=-\frac{B_{axis}R_0 y}{x^2+y^2}-\frac{B_{axis}xz}{q(x^2+y^2)}
\eqno{(30.a)}
$$
$$
B_y=B_{\phi} \text{cos}\phi-B_{\theta} \text{sin}\theta \text{sin}\phi=\frac{B_{axis}R_0 x}{x^2+y^2}-\frac{B_{axis}yz}{q(x^2+y^2)}
\eqno{(30.b)}
$$
$$
B_z=B_{\theta} \text{cos}\theta=\frac{B_{axis}(\sqrt{x^2+y^2}-R_0)}{q\sqrt{x^2+y^2}}
\eqno{(30.c)}
$$
 Starting with the initial position $\vec r_0=(R_0+0.25a,0,0)^T=(1.82m,0,0)^T$ and the initial velocity $\vec v_0=(0,2\times 10^4 m/s,2\times 10^5 m/s)$, the projection of the particle’s trajectory on the  plane forms a closed banana orbit, and it will transform into a transit orbit when the initial velocity is changed into $\vec v_0=(0,8\times 10^4 m/s,2\times 10^5 m/s)$. The Boris algorithm and $G_h^2$ are implemented with relatively large time step size $\omega_{c0} \Delta t=0.1$, and the numerical results of the banana and transit orbit are shown in Figure \ref{fig 1} and Figure \ref{fig 2} respectively. The time integration interval is $[0,T_0],\omega_{c0} T_0 =2.54\times 10^4$ for the banana orbit in Figure \ref{fig 1} and $[0,T_1],\omega_{c0} T_1=1.38\times 10^4$ for the transit orbit in Figure \ref{fig 2}. Both algorithms have correctly achieved the trajectory of the trapped and transit particle due to their volume-preserving nature. 

\begin{figure}
\centering
\begin{subfigure}[b]{0.49\textwidth}
  \includegraphics[width=\textwidth]{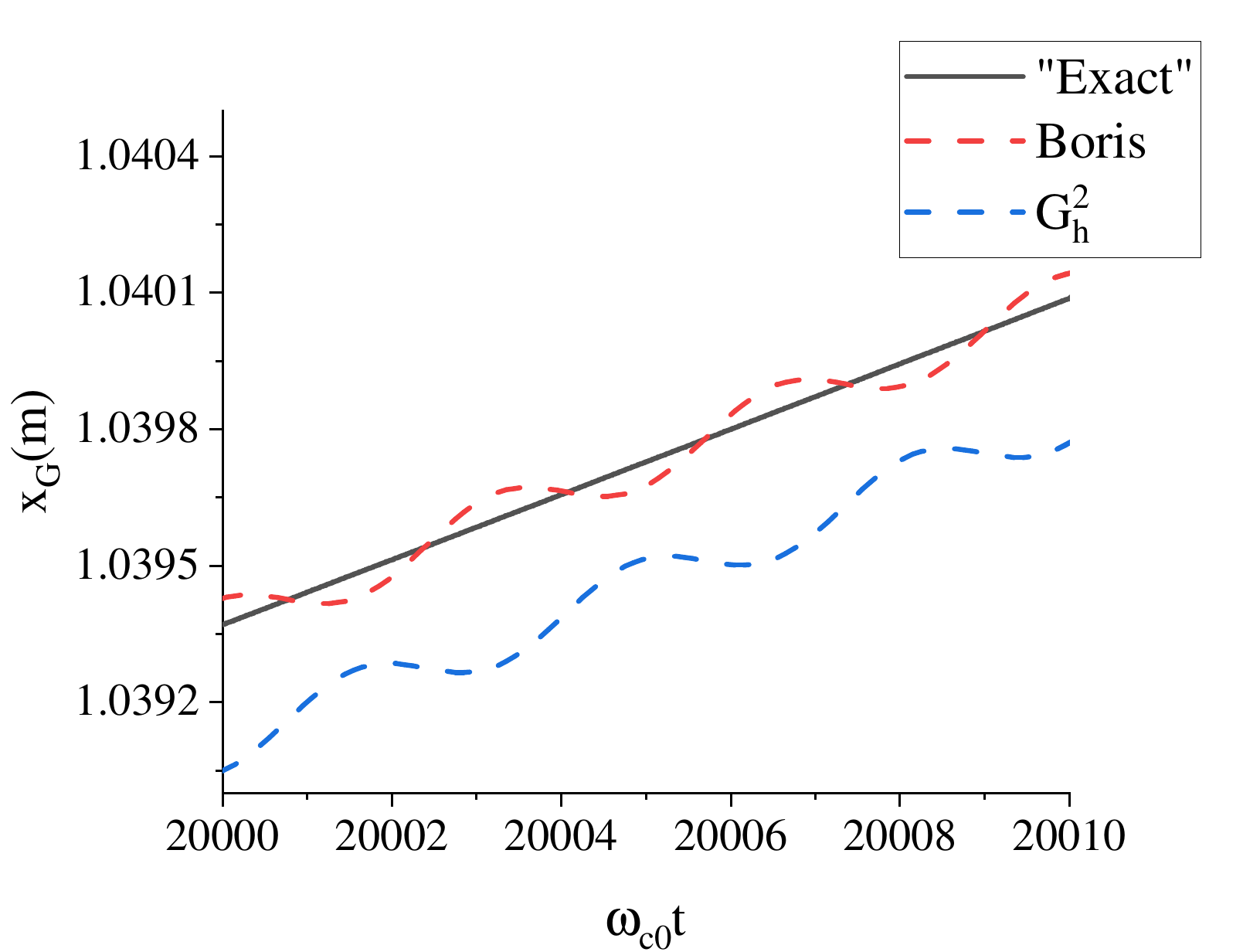}
  \caption{$x_G$ in $[20000,20010]$}
  \label{fig 3-a}
\end{subfigure}
\hfill 
\begin{subfigure}[b]{0.49\textwidth}
  \includegraphics[width=\textwidth]{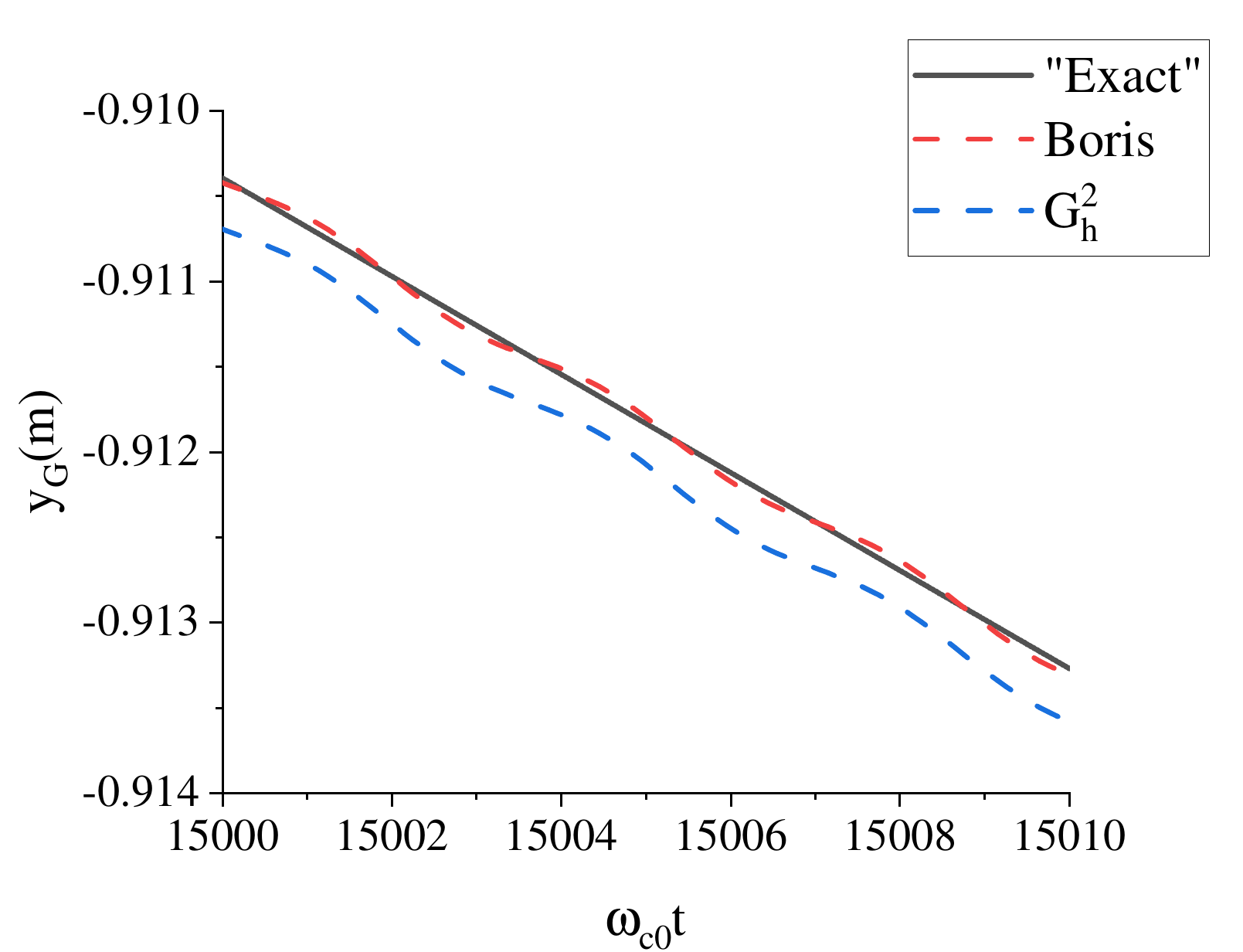}
  \caption{$y_G$ in $[15000,15010]$}
  \label{fig 3-b}
\end{subfigure}
\hfill 
\begin{subfigure}[b]{0.49\textwidth}
  \includegraphics[width=\textwidth]{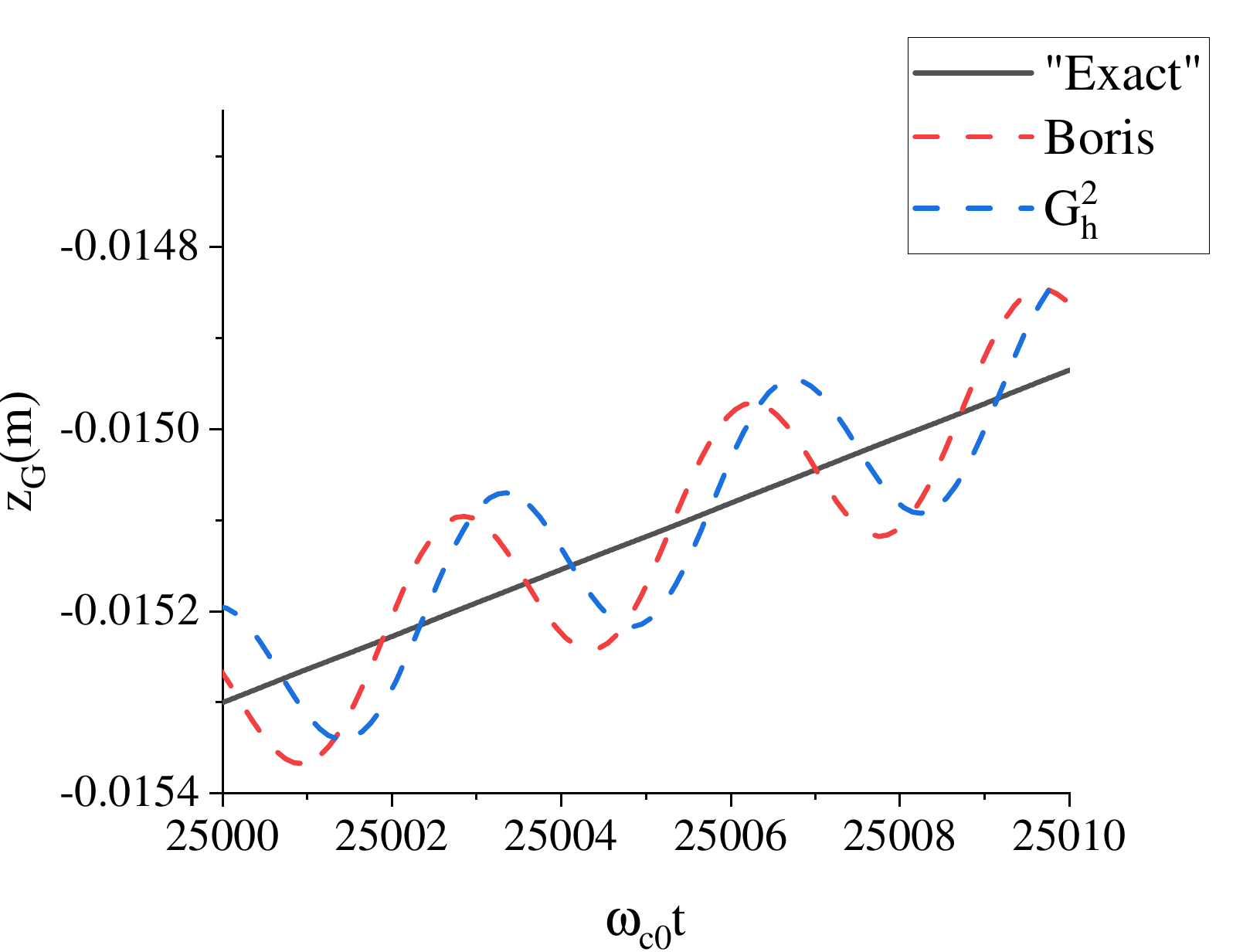}
  \caption{$z_G$ in $[25000,25010]$}
  \label{fig 3-c}
\end{subfigure}

\caption{Time-dependent numerical and “exact” results of the slow-scale motions $\vec r_G$ in certain time intervals. The red dashed lines (the Boris algorithm) oscillate closely around the black solid lines (“exact” solutions) in each graph. And the blue dashed lines ($G_h^2$) are significantly diverged from the black solid lines in (a) and (b). }
\label{fig 3}
\end{figure}

\begin{figure}
\centering
\begin{subfigure}[b]{0.49\textwidth}
  \includegraphics[width=\textwidth]{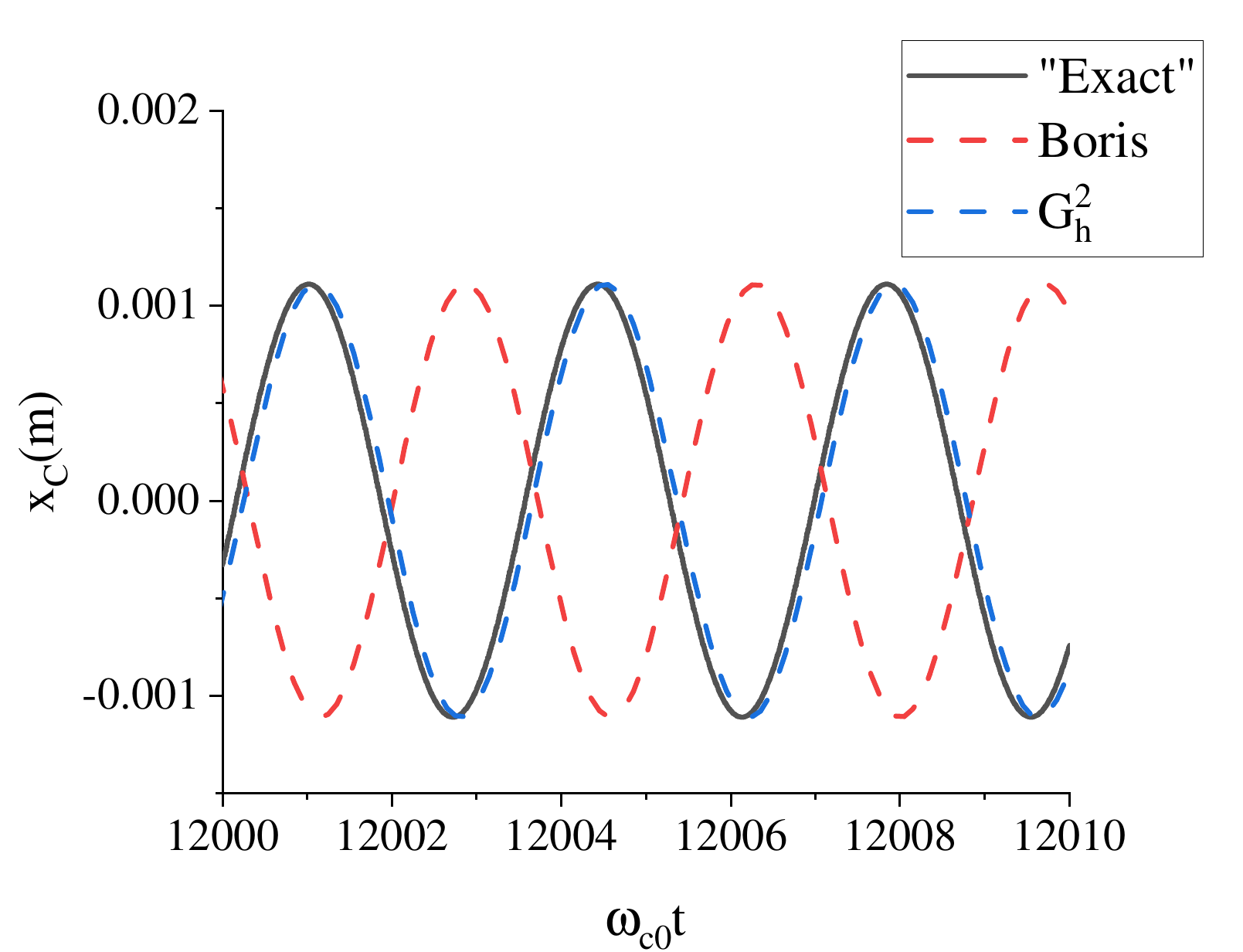}
  \caption{$x_C$ in $[12000,12010]$}
  \label{fig 4-a}
\end{subfigure}
\hfill 
\begin{subfigure}[b]{0.49\textwidth}
  \includegraphics[width=\textwidth]{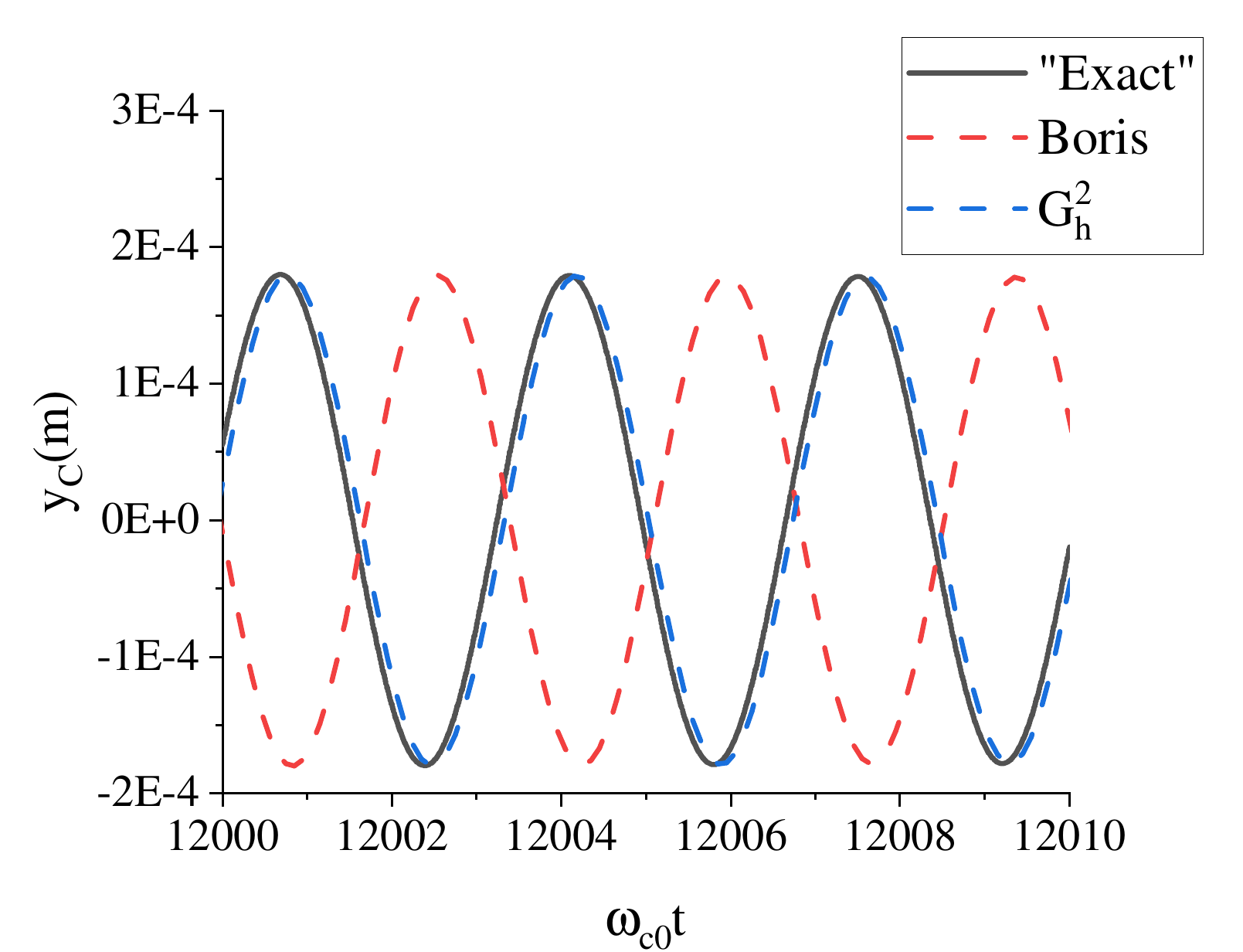}
  \caption{$y_C$ in $[12000,12010]$}
  \label{fig 4-b}
\end{subfigure}
\hfill 
\begin{subfigure}[b]{0.49\textwidth}
  \includegraphics[width=\textwidth]{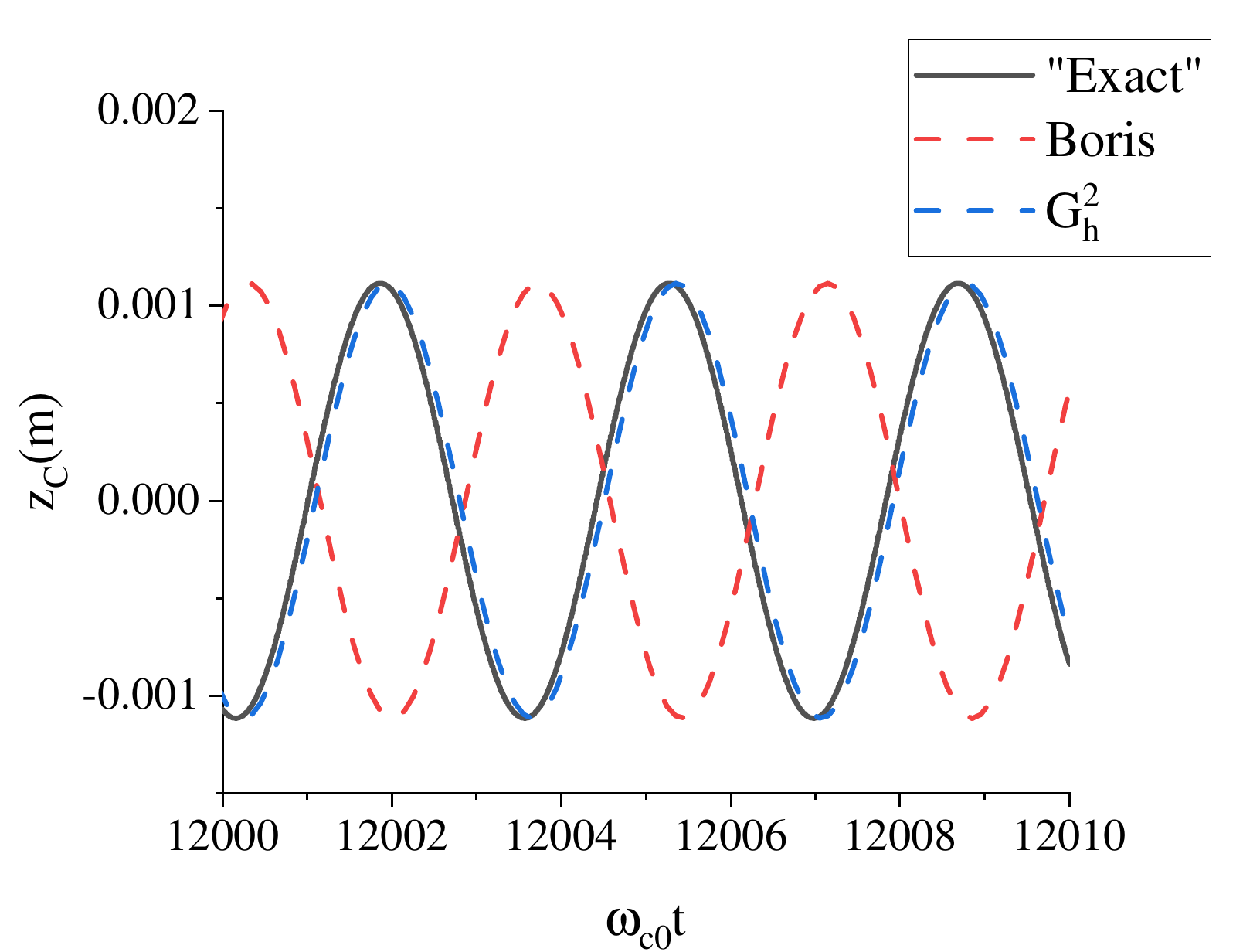}
  \caption{$z_C$ in $[12000,12010]$}
  \label{fig 4-c}
\end{subfigure}

\caption{Time-dependent numerical and “exact” results of the fast-scale motions $\vec r_C$ in certain time intervals. The blue dashed lines ($G_h^2$) almost overlap with the black solid lines (“exact” solutions), while the red dashed lines (the Boris algorithm) accumulate visible phase errors in each graph. }
\label{fig 4}
\end{figure}

Now we analyze their differences by dividing the numerical solutions of position $\vec r$ into two components: the slow-scale guiding center motions $\vec r_G$ and the fast-scale cyclotron motions $\vec r_C$
$$
\vec r_C=-\frac{m_i}{e} \frac{\vec v \times \vec B(\vec r)}{B^2(\vec r)}
\eqno{(31.a)}
$$
$$
\vec r_G=\vec r-\vec r_C=\vec r+\frac{m_i}{e} \frac{\vec v \times \vec B(\vec r)}{B^2(\vec r)}
\eqno{(31.b)}
$$
Shown in Figure \ref{fig 3} and Figure \ref{fig 4} is the separated time-dependent numerical results of the banana orbit by both algorithms with $\omega_{c0} \Delta t=0.1$ in selected time intervals of equal length of 10, together with the 'exact' solutions obtained by an extremely minuscule time step ($\omega_{c0} \Delta t=10^{-4}$, which is virtually impossible to achieve in numerical simulations, and both algorithms generate identical results in this case. Here we select the result derived by the Boris algorithm). In Figure \ref{fig 3} we display the numerical and 'exact' results of the slow-scale motions $\vec r_G$. The Boris algorithm closely matches the analytical solution for slow-scale motions. In contrast, $G_h^2$ produces noticeable discrepancies, which are particularly evident in Figure \ref{fig 3-a} and Figure \ref{fig 3-b}. Regarding the instances of the fast-scale motions $\vec r_C$ depicted in Figure \ref{fig 4}, the performances of the two algorithms are diametrically opposed: solutions derived from $G_h^2$ almost perfectly overlap with the analytical solution, and solutions of the Boris algorithm exhibit a significant phase error. 

\begin{figure}
\centering
\begin{subfigure}[b]{0.49\textwidth}
  \includegraphics[width=\textwidth]{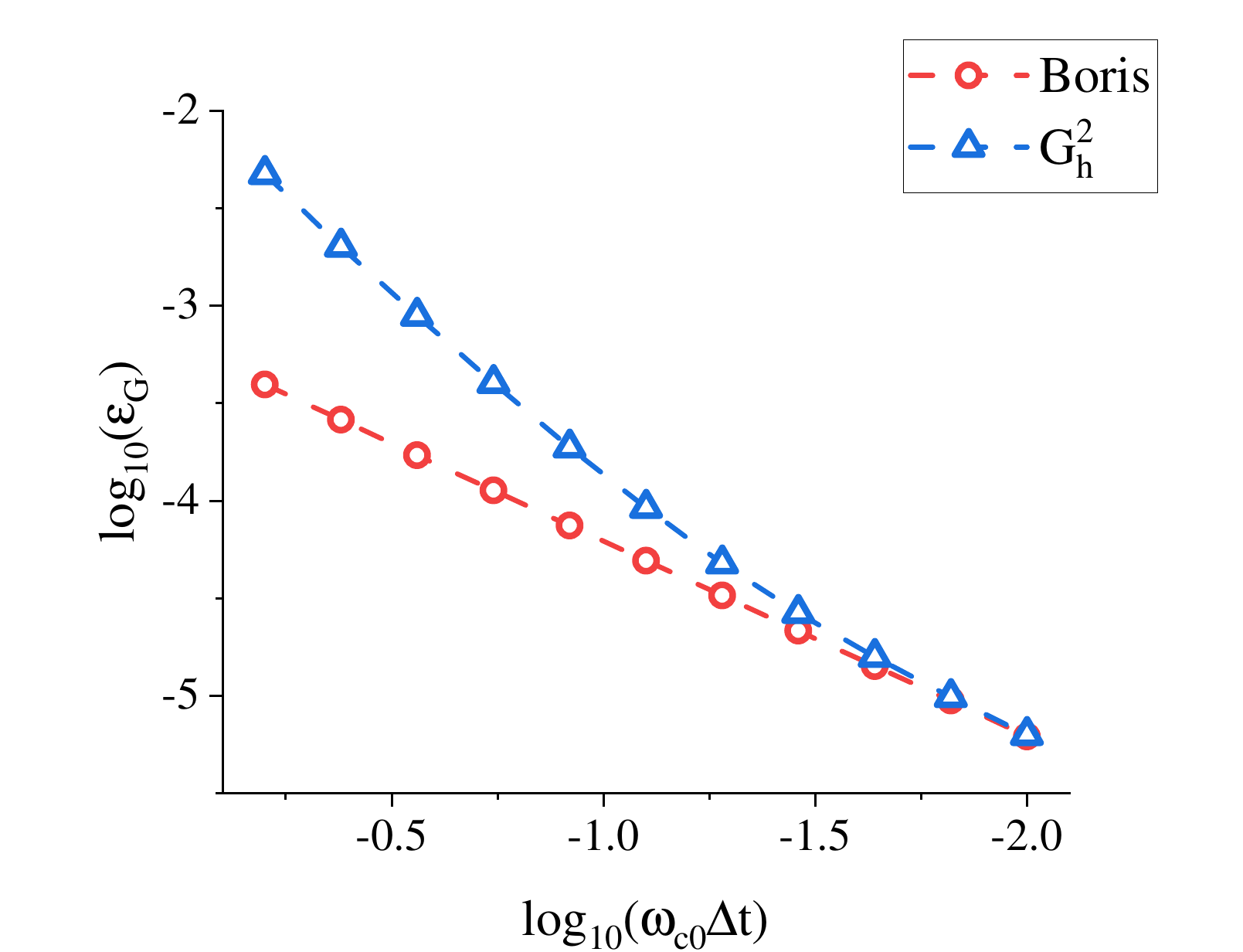}
  \caption{$\epsilon_G$}
  \label{fig 5-a}
\end{subfigure}
\hfill 
\begin{subfigure}[b]{0.49\textwidth}
  \includegraphics[width=\textwidth]{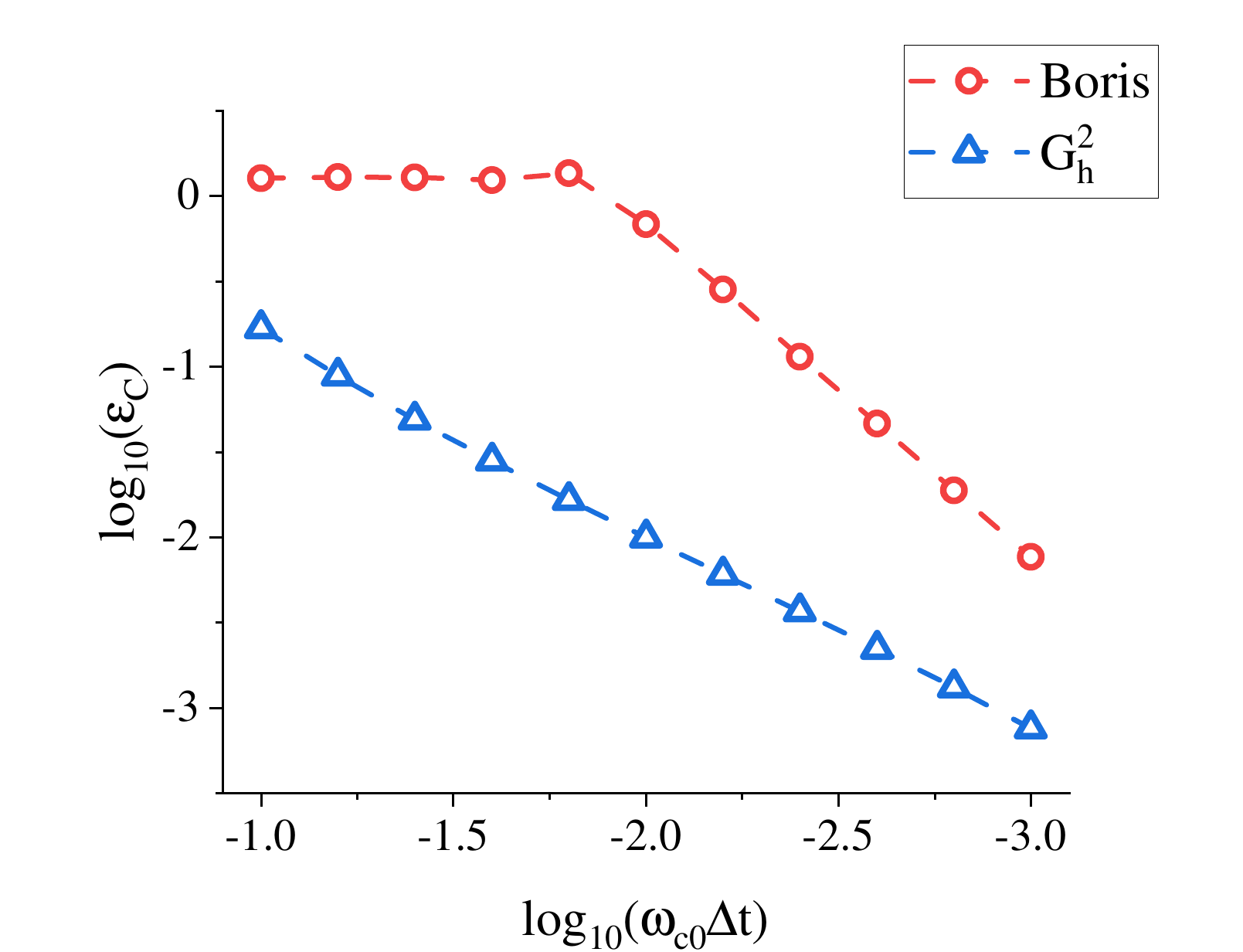}
  \caption{$\epsilon_C$}
  \label{fig 5-b}
\end{subfigure}

\caption{Global relative errors of $\vec r_G$ and $\vec r_C$ as a function of time step size $\Delta t$ by both algorithms.}
\label{fig 5}
\end{figure}

To test the efficiency of the two algorithms in handling diverse scales of motions, Figure \ref{fig 5} compares the global relative errors over the entire time integration interval $[0,T_0]$ of $\vec r_G$ and $\vec r_C$ as a function of time step size $\Delta t$. Here, the relative error is defined by
$$
\epsilon_G = \frac{1}{N} \sum_{m=0}^{N-1} \sqrt{\frac{|\vec r_G^{"exact"}(t_m)-\vec r_G^{numerical}(t_m)|^2}{|\vec r_G^{"exact"}(t_m)|^2}}
\eqno{(32.a)}
$$
$$
\epsilon_C = \frac{1}{N} \sum_{m=0}^{N-1} \sqrt{\frac{|\vec r_C^{"exact"}(t_m)-\vec r_C^{numerical}(t_m)|^2}{|\vec r_C^{"exact"}(t_m)|^2}}
\eqno{(32.b)}
$$
with $N=\frac{T_0}{\Delta t}$ the total number of time grids, and $t_m=(m+\frac{1}{2})\Delta t$ the corresponding time of the position variables in the $m-th$ time step. It can be observed from Figure \ref{fig 5-a} that the relative error of Boris method converges faster in the case of slow-scale motions, while $G_h^2$ is more efficient in the fast-scale motions as shown in Figure \ref{fig 5-b}. The outcomes correspond to the phenomena illustrated in Figure \ref{fig 3} and Figure \ref{fig 4} respectively.

\begin{figure}
\centering
\begin{subfigure}[b]{0.49\textwidth}
  \includegraphics[width=\textwidth]{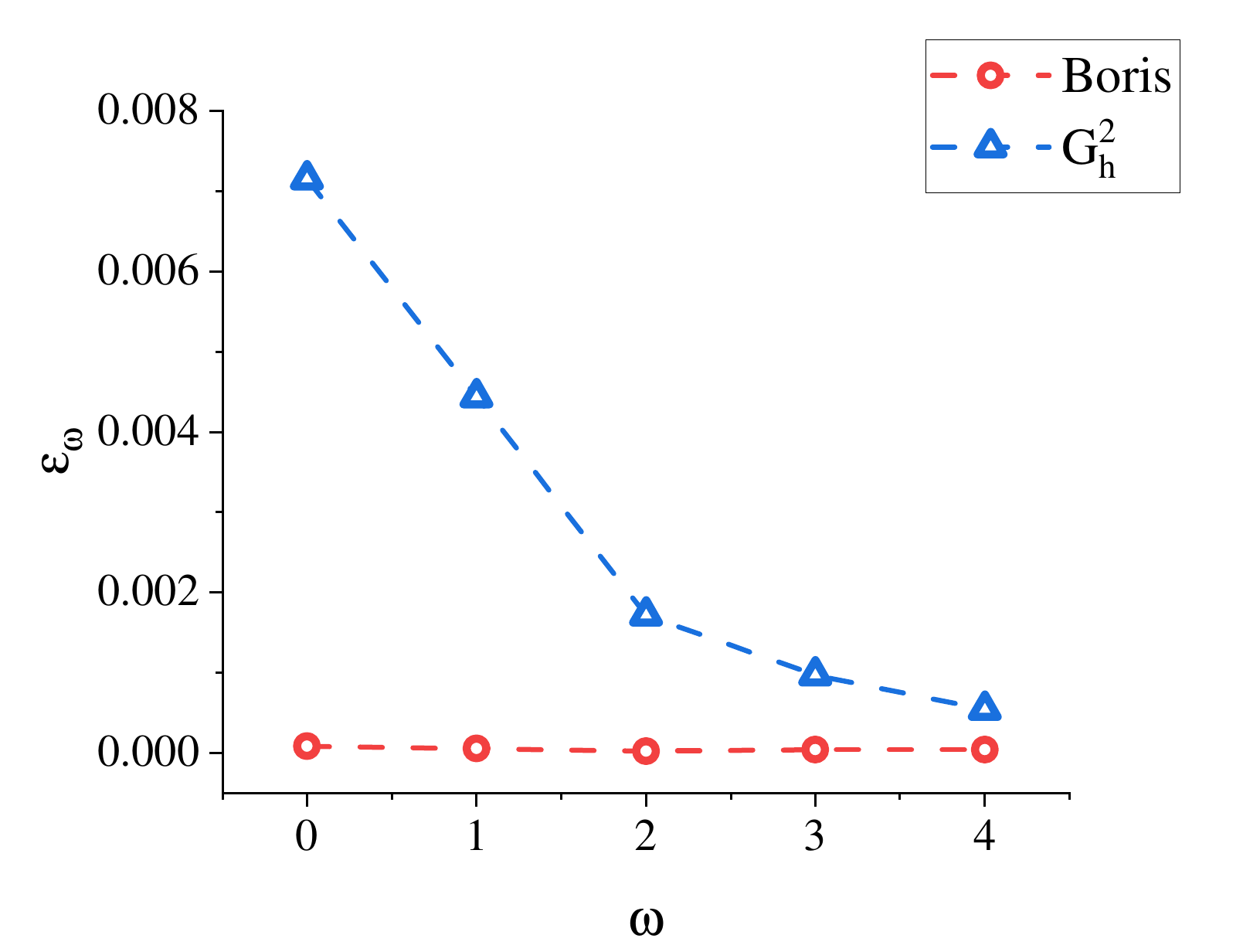}
  \caption{$\epsilon_{\omega}$ in $[0,4]$}
  \label{fig 6-a}
\end{subfigure}
\hfill 
\begin{subfigure}[b]{0.49\textwidth}
  \includegraphics[width=\textwidth]{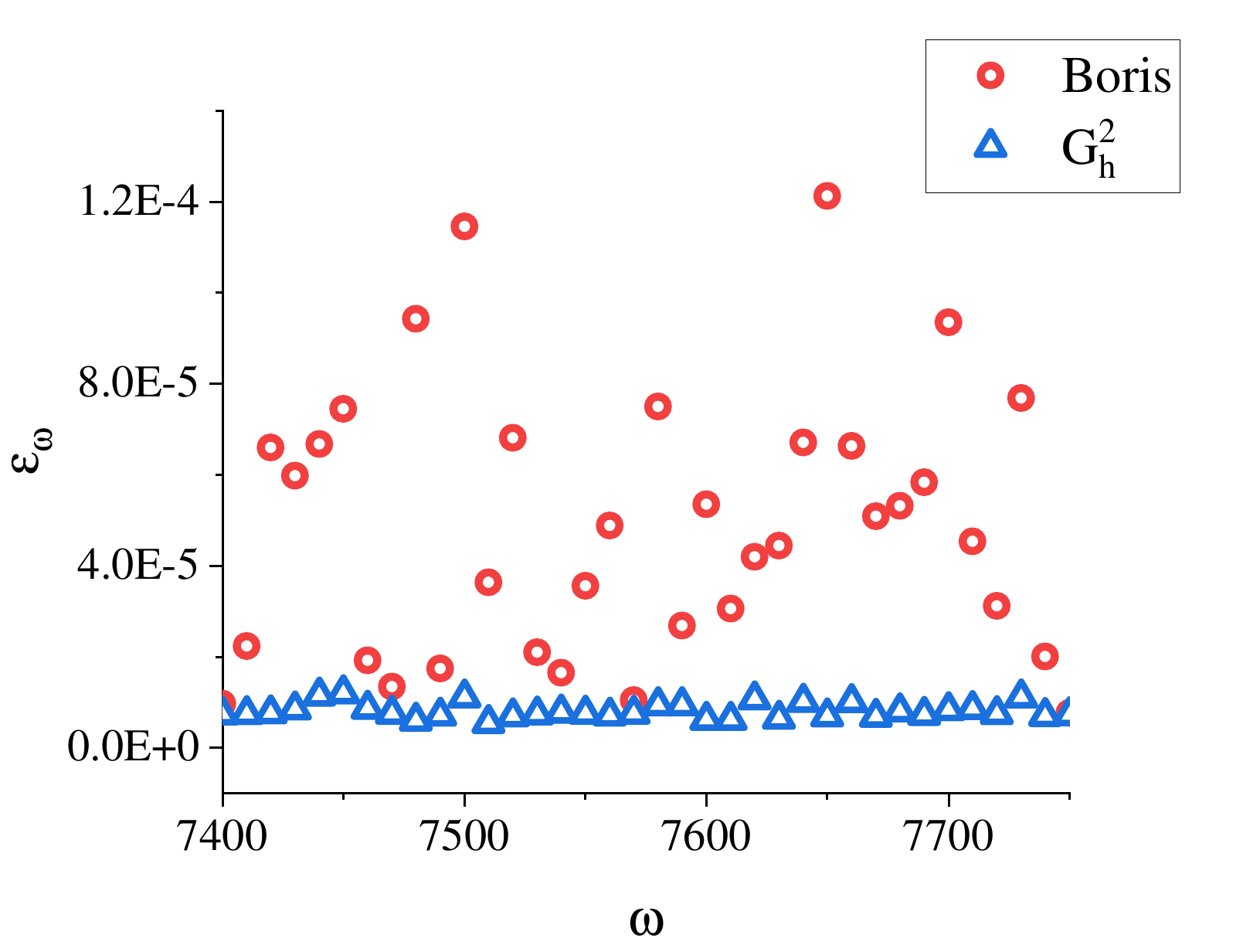}
  \caption{$\epsilon_{\omega}$ in $[7400,7750]$}
  \label{fig 6-b}
\end{subfigure}

\caption{2-norm errors of DFT coefficients $\epsilon_{\omega}$ for various scales of $\omega$.(a).small $\omega$ related to slow-scale guiding-center motions.(b).large $\omega$ related to fast-scale cyclotron motions}
\label{fig 6}
\end{figure}

Now we compare the convergence rate of the DFT coefficients given by equation (13), which is measured by the 2-norm of errors between the numerical solutions and the "exact" solutions
$$
\epsilon_{\omega}=||\vec f^{"exact"}(\omega)-\vec f^{numerical}(\omega)||_2
\eqno{(33)}
$$
Figure \ref{fig 6-a} shows the results of small $\omega$( i.e. $\omega \in [0,4] \sim O(1)$ , which reflects the slow-scale guiding center motions). The time step size is $\omega_{c0} \Delta t=10^{-0.2}$ which is the largest size in Figure \ref{fig 5-a}. The outcomes of the Boris algorithm remain consistently similar to the "exact" cases while deviations are observed in the situations of $G_h^2$. In the case of large $\omega$ (i.e. $\frac{2\pi\omega}{\omega_{c0} T}=\frac{B}{B_0}\sim O(1)$, which reflects the fast-scale cyclotron motions), the magnetic field strength $B$ ranges from 1.84T to 1.92T over the entire banana period, resulting in non-trivial values of $\vec f(\omega)$ within the approximate range of $[7400,7750]$. In Figure \ref{fig 6-b} we display the numerical errors with time step size $\omega_{c0} \Delta t=0.1$ (which is the largest size in Figure \ref{fig 5-b}) and $\omega \in [7400,7750]$, and the situation starkly contrasts with Figure \ref{fig 6-a}: the Boris algorithm yields completely distinct results to the “exact” solutions, with only minor errors occurring as for $G_h^2$. 

The above experiments corroborates the theoretical analysis of the preceding section: the guiding center orbit (slow-scale motions, low-frequency component of position) become more accurate utilizing the Boris algorithm, while the gyro-motion (fast-scale motions, high-frequency component of position) is better described by $G_h^2$ due to the different convergence rate of the corresponding DFT coefficients.

\begin{figure}
\centering
\begin{subfigure}[b]{0.49\textwidth}
  \includegraphics[width=\textwidth]{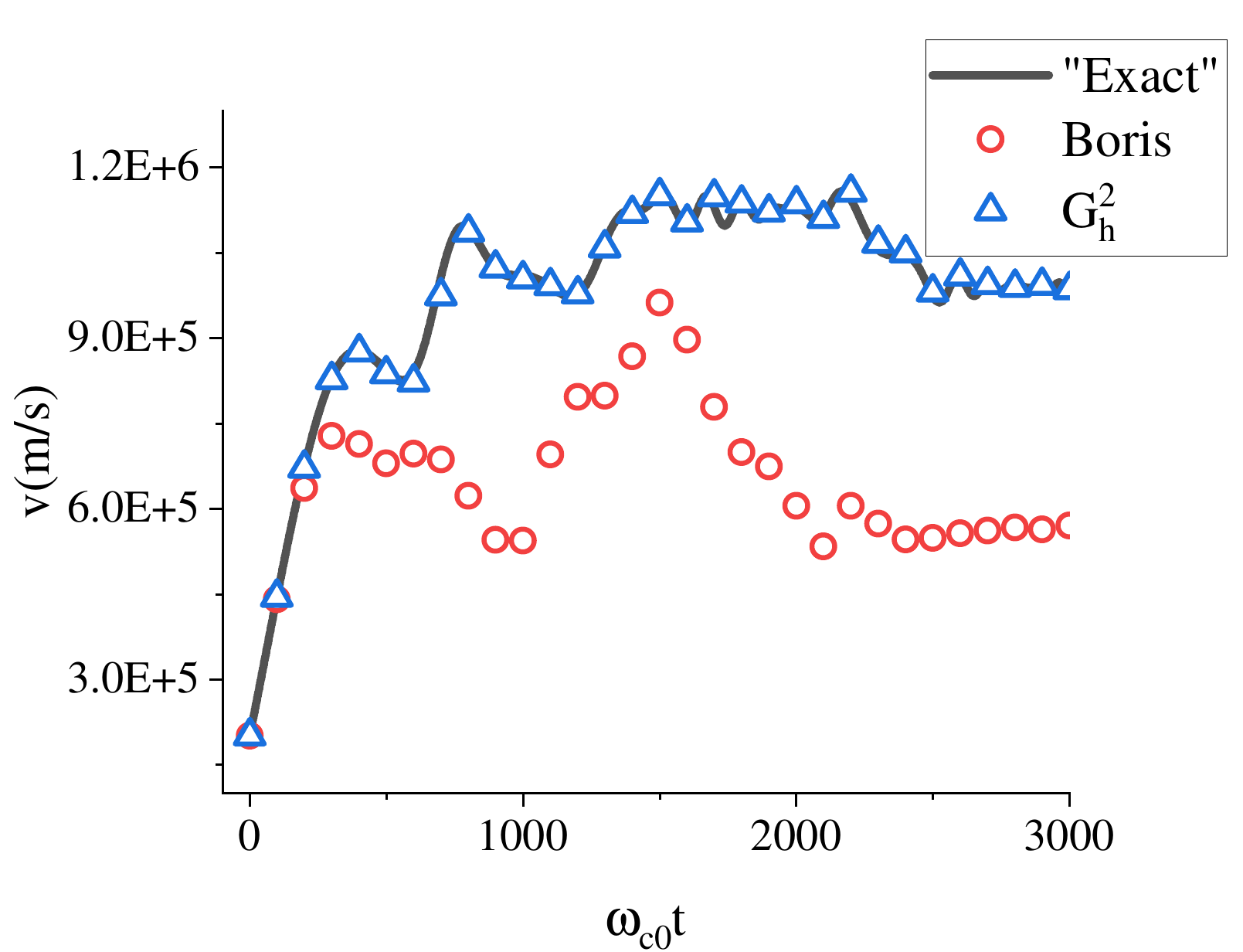}
  \caption{$v$ in $[0,3000]$}
  \label{fig 7-a}
\end{subfigure}
\hfill 
\begin{subfigure}[b]{0.49\textwidth}
  \includegraphics[width=\textwidth]{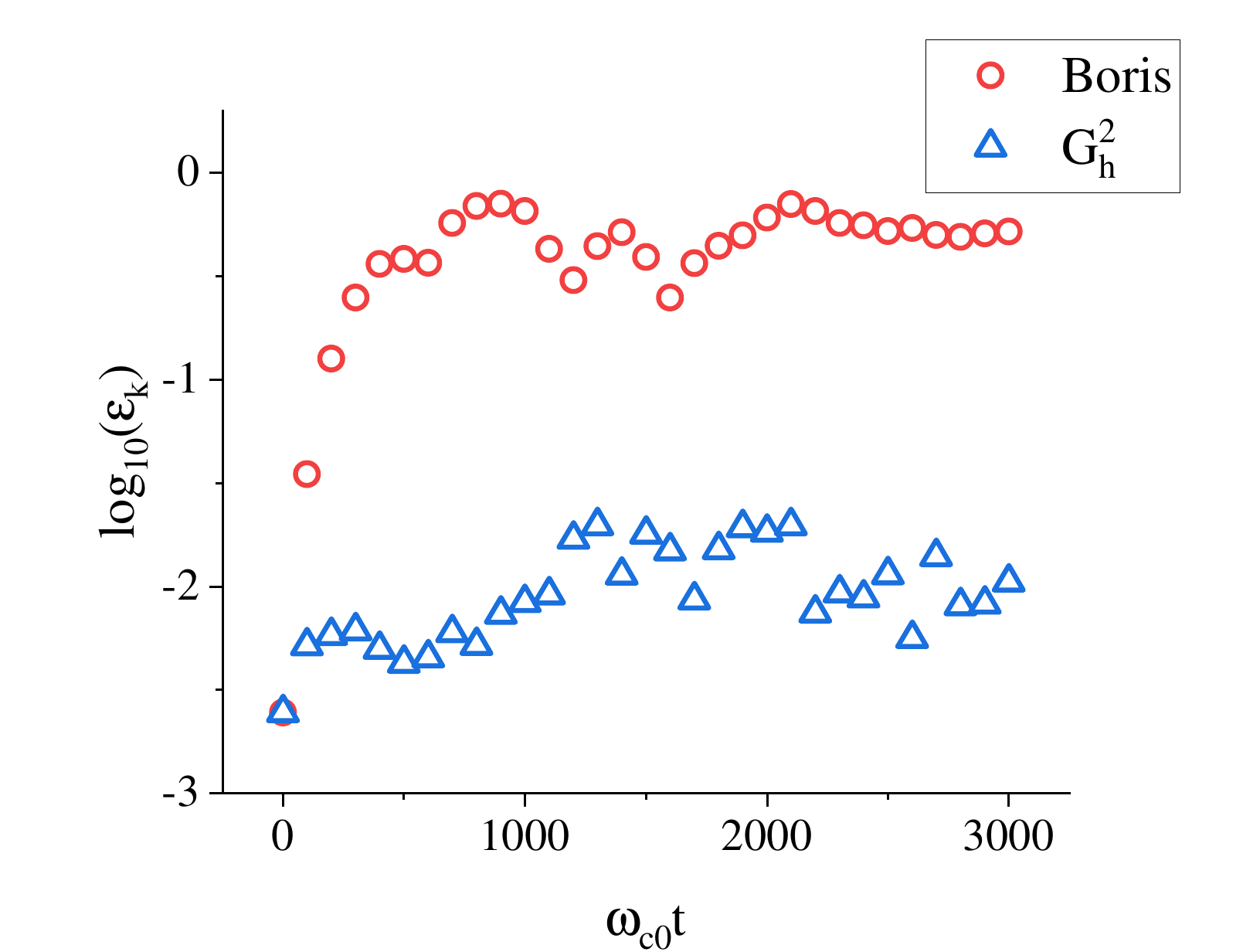}
  \caption{$\epsilon$ in $[0,3000]$}
  \label{fig 7-b}
\end{subfigure}

\caption{Numerical results of the trapped particle with an additional resonant electric field at the gyro-frequency and fixed time step size $\omega_{c0}\Delta t=0.1$. The time integration interval is $[0,T_2],\omega_{c0}T_2=3 \times 10^3$. (a). velocity magnitude $v$ as a function of time. (b). relative error of kinetic energy $\epsilon$. The numerical errors of the Boris algorithm is much larger than $G_h^2$ in this problem.}
\label{fig 7}
\end{figure}

\begin{figure}
\centering
\begin{subfigure}[b]{0.49\textwidth}
  \includegraphics[width=\textwidth]{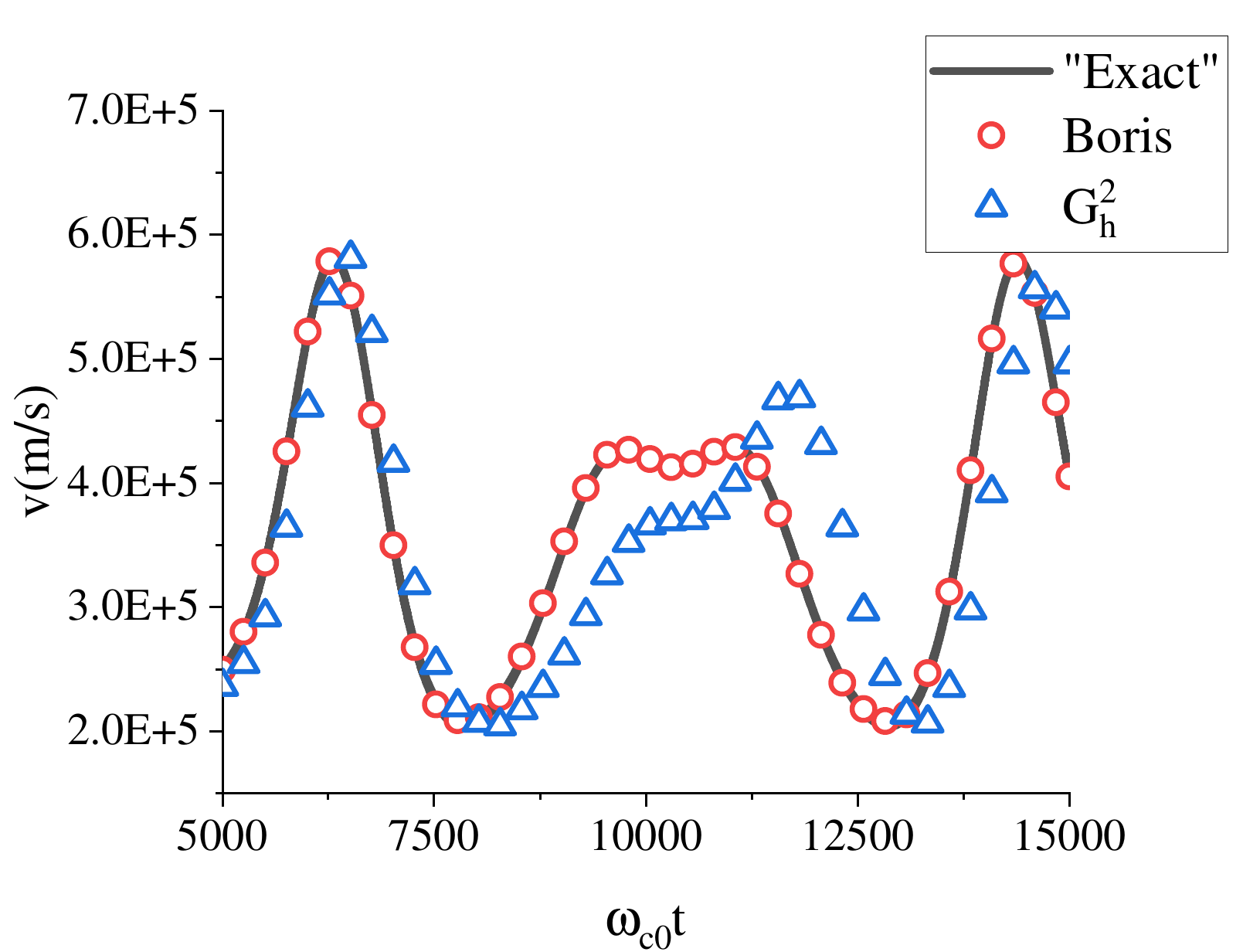}
  \caption{$v$ in $[5000,15000]$}
  \label{fig 8-a}
\end{subfigure}
\hfill 
\begin{subfigure}[b]{0.49\textwidth}
  \includegraphics[width=\textwidth]{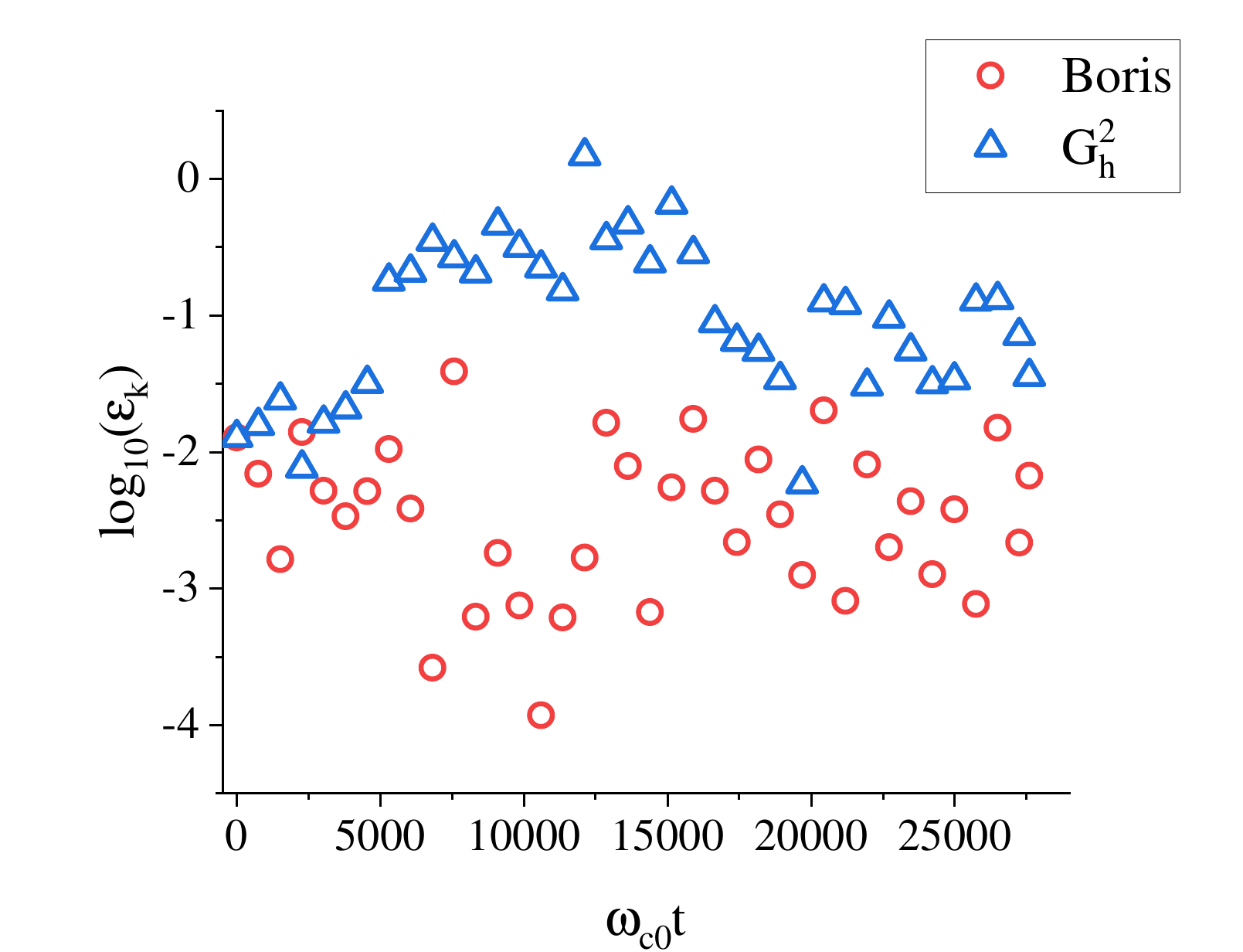}
  \caption{$\epsilon$ in in $[0,28000]$}
  \label{fig 8-b}
\end{subfigure}

\caption{Numerical results of the transit particle with an additional resonant electric field at the transit frequency $\frac{2\pi}{T_1},\omega_{c0} T_1=1.38\times 10^4$ and fixed time step size $\omega_{c0}\Delta t=10^{-0.2}$. The time integration interval is $[0,2T_1]$. (a). velocity magnitude $v$ in [5000,15000] as a function of time. (b). relative error of kinetic energy $\epsilon$. The numerical errors of $G_h^2$ is much larger than the Boris algorithm in this problem.}
\label{fig 8}
\end{figure}

Finally, we examine the process of wave heating by introducing a resonant electric field on the z-axis $\vec E=(0,0,E_0cos(\omega_0t)),E_0=5 \times 10^3 V/m$  with frequency $\omega_0$. The velocity magnitude $v_k$ and relative error of kinetic energy $\epsilon_k$ in the $k-th$ time step are calculated to quantify the impact of heating 
$$
v_k=|\vec v_k|,\epsilon_k=\frac{(v_k^{"Exact"})^2-(v_k^{Numerical})^2}{(v_k^{"Exact"})^2}
\eqno{(34)}
$$
Firstly, we consider the trapped particle with the same initial condition of banana orbit, and the frequency of the electric field $\omega_0$ is set to be $\omega_0=\frac{B(\vec r)}{B_0} \omega_{c0}$ to match the gyro-frequency. The simulation over $[0,T_2]$ with $\omega _{c0} T_2=3\times 10^3$ and fixed time step size $\omega_{c0} \Delta t=0.1$ is shown in Figure \ref{fig 7}. In this case, $G_h^2$ provides superior numerical results compared to the Boris algorithm due to its higher accuracy of cyclotron motions. And shown in Figure \ref{fig 8} is the result of the transit particle with a low-frequency resonant electric field $\omega_{0}=\frac{2\pi}{T_1}$ with $\omega_{c0}T_1=1.38\times 10^4$ to match the transit period. The time step size is $\omega_{c0} \Delta t=10^{-0.2}$, and the time integration interval is $[0,2T_1]$. It is noticed that the Boris algorithm handles the low-frequency electric field better than $G_h^2$ in this example.

\begin{figure}
\centering
\begin{subfigure}[b]{0.49\textwidth}
  \includegraphics[width=\textwidth]{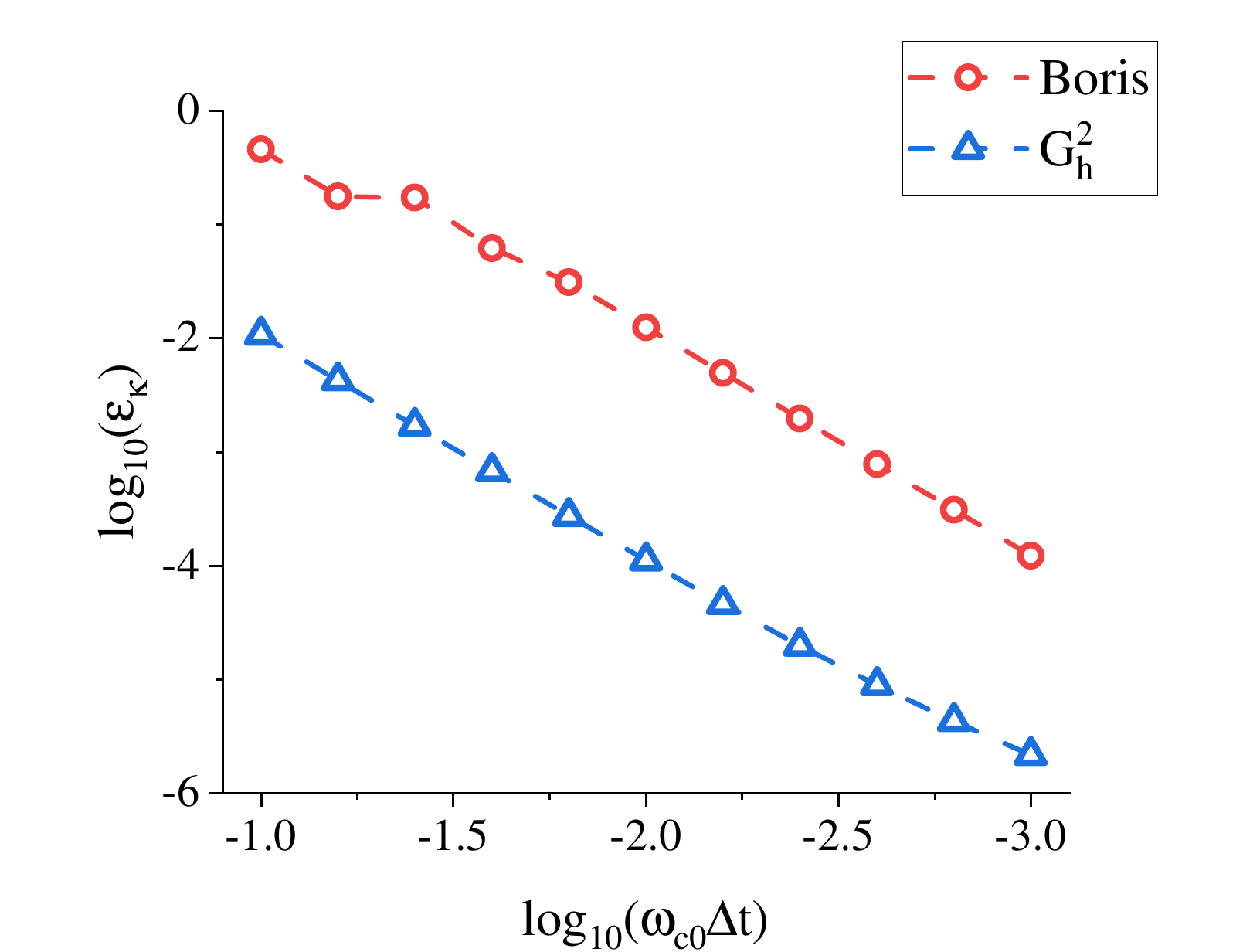}
  \caption{}
  \label{fig 9-a}
\end{subfigure}
\hfill 
\begin{subfigure}[b]{0.49\textwidth}
  \includegraphics[width=\textwidth]{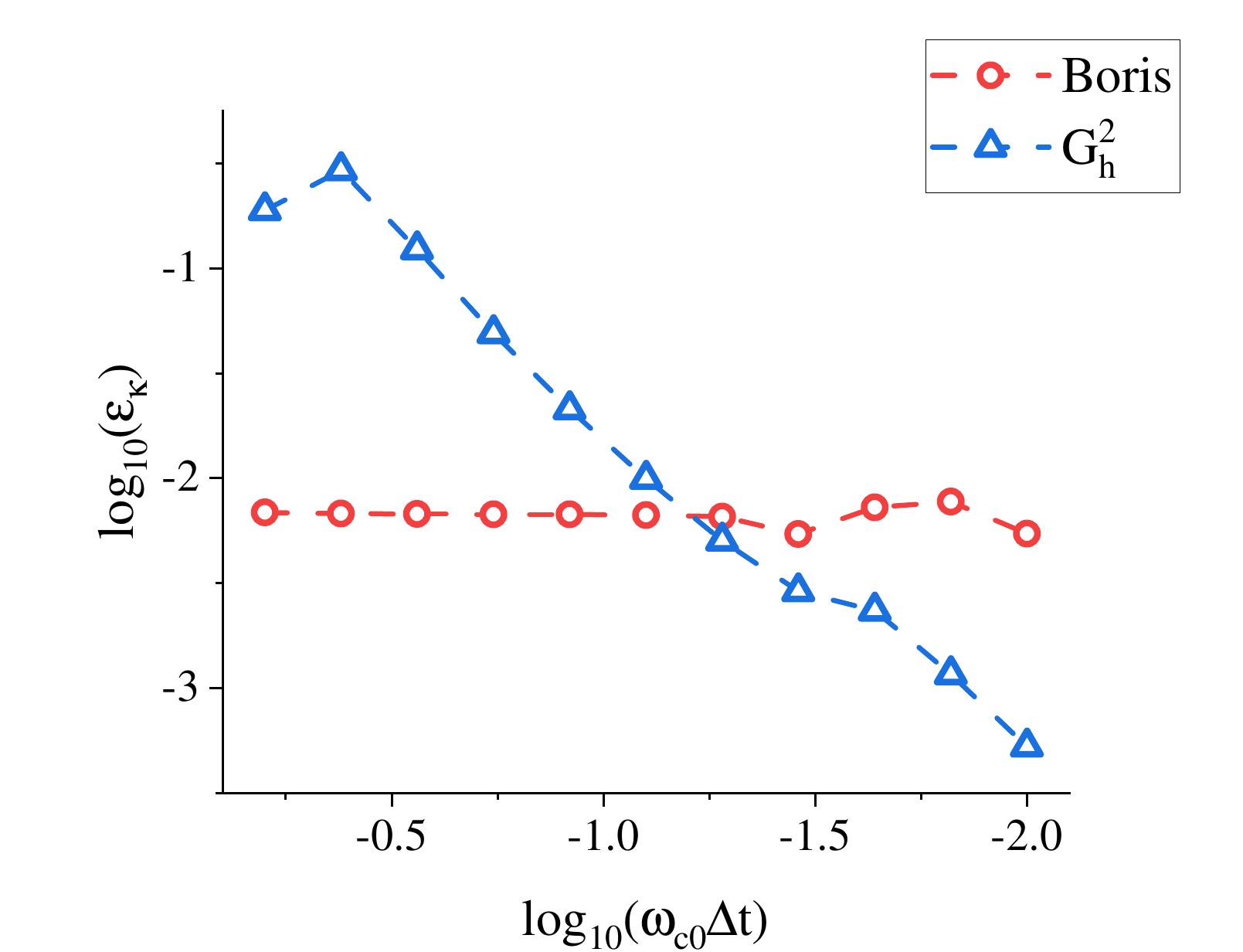}
  \caption{}
  \label{fig 9-b}
\end{subfigure}

\caption{Global relative errors of kinetic energy $\epsilon_{\kappa}$ as a function of time step size $\Delta t$ by both algorithms. (a). trapped particle with high-frequency resonant electric field. (b). transit particle with low-frequency resonant electric field.}
\label{fig 9}
\end{figure}

To analyze further, in Figure \ref{fig 9} we compute the global relative error of kinetic energy as a function of time step size $\Delta t$. Here, the global relative error $\epsilon_{\kappa}$ is defined by
$$
\epsilon_{\kappa}=\frac{1}{N} \sum_{k=1}^N \epsilon_k
\eqno{(35)}
$$
with $N=\frac{T_2}{\Delta t}$ or $N=\frac{2T_1}{\Delta t}$ the total number of time grids of the above two cases. For the case of high-frequency electric field in Figure \ref{fig 9-a}, both algorithms converge as the time step size diminished to zero, and $G_h^2$ allows for a significant larger time step size than the Boris algorithm (approximately one order of magnitude) to reach the same level of accuracy. The situation of low-frequency electric field in Figure \ref{fig 9-b} is slightly different. The precision of the Boris algorithm is virtually unaltered in the depicted range, while $G_h^2$  maintains the convergent numerical results. Nonetheless, the Boris algorithm still possesses substantial advantages at large time step sizes. One can observe that the two algorithms exhibit distinct advantages when addressing issues with varying characteristic frequencies, highlighting the importance of selecting the appropriate numerical scheme based on the specific motion scales or characteristic frequencies of interest.

\section{Conclusion}
~~~~In this study, we have conducted a comprehensive analysis of the efficacy of various volume-preserving algorithms in accurate single particle orbit simulations, particularly in context of the phase stability across various frequencies. Our findings, corroborated by both theoretical analysis and numerical experiments, consistently indicate that the Boris algorithm possesses superiority in simulating slow-scale guiding center motions, rendering it the optimal choice for physical problems characterized by low frequency scale, while $G_h^2$ have demonstrated enhanced efficiency in the simulation of fast-scale gyro-motions and high characteristic frequency. It is rather challenging to accurately calculate the guiding-center motion and gyro-motion simultaneously within the constraints of limited computational resources. Thus, the selection of the appropriate numerical scheme is pivotal and should be informed by the characteristic frequencies in specific physical problems of interest. For instance, for low frequency wave such as drift-wave turbulence and shear Alfvén waves,  the Boris algorithm is typically the more favorable option. Conversely, for problems characterized by briefer time scales and high frequencies, such as RF heating, high frequency turbulence and ICE, $G_h^2$ may yield superior results.

\section*{Appendix: Approximation}
~~~~We describe here the specific derivation process of equations (19) under the condition of approximation given by equation (16).  Substituting equation (17.b) into (18) yields
$$
\vec f(\omega)=\frac{\Delta t}{T}\text{exp}\left(-\frac{1}{2}\Delta t \cdot i\right)(I_1+I_2)
\eqno{(A-1.a)}
$$
with 
$$
I_1=\left(\vec r_0+\frac{1}{2}\vec v_0\Delta t\right)\sum_{m=0}^{N-1} \text{exp}(-km\Delta t\cdot i)
\eqno{(A-1.b)}
$$
and
$$
I_2=\sum_{m=1}^{N-1} \sum_{j=1}^{m} \text{exp}(-km\Delta t \cdot i)\vec v_j\Delta t
\eqno{(A-1.c)}
$$
Since $\sum_{m=0}^{N-1} \text{exp}(-km\Delta t\cdot i)=\frac{1-\text{exp}(-kN\Delta t \cdot i)}{1-\text{exp}(-k\Delta t\cdot i)}=\frac{1-\text{exp}(-2\pi\omega \cdot i)}{1-\text{exp}(-k\Delta t\cdot i)}=0$, we have $I_1=0$.  The simplification of $I_2$ is derived by exchanging the summation sequence
$$
I_2=\sum_{j=1}^{N-1} \sum_{m=j}^{N-1} \text{exp}(-km\Delta t\cdot i)\vec v_j \Delta t
=\frac{\Delta t}{1-\text{exp}(-k\Delta t \cdot i)}\left(\sum_{j=1}^{N-1}\text{exp}(-kj\Delta t\cdot i)\vec v_j-\sum_{j=1}^{N-1}\vec v_j\right)
\eqno{(A-2)}
$$
By inserting the explicit formulation (17.a) for $\vec v_j$ into the preceding equation, we derive the exact expressions for $\lambda_1$, $\lambda_2$, and $\lambda_3$
$$
\lambda_1=-N\Delta t =-T
\eqno{(A-3.a)}
$$
$$
\lambda_2=\left(\sum_{j=1}^{N-1} \prod_{n=1}^{j}  \mu_{2,n} - \sum_{j=1}^{N-1} \prod_{n=1}^{j}  \lambda_{2,n}\right)\Delta t
\eqno{(A-3.b)}
$$
$$
\lambda_3=\left(\sum_{j=1}^{N-1} \prod_{n=1}^{j}  \mu_{3,n} - \sum_{j=1}^{N-1} \prod_{n=1}^{j}  \lambda_{3,n}\right)\Delta t
\eqno{(A-3.c)}
$$
with
$$
\mu_{2,n}=\lambda_{2,n} \text{exp}(-k\Delta t\cdot i),\mu_{3,n}=\lambda_{3,n} \text{exp}(-k\Delta t\cdot i)
\eqno{(A-3.d)}
$$
    Now we are in the position to conduct approximations. We focus our analysis solely on $\lambda_2$; an identical process can be similarly applied to $\lambda_3$, yielding analogous results.  We  henceforth substitute all subscripts $n$ with $N-n$ 
$$
\lambda_2=\left(\sum_{j=1}^{N-1} \prod_{n=1}^{j}  \mu_{2,N-n} - \sum_{j=1}^{N-1} \prod_{n=1}^{j}  \lambda_{2,N-n}\right)\Delta t
\eqno{(A-4)}
$$
This will facilitate the formulation of $\lambda_2$ in a recursive manner
$$
a_n=(1+a_{n-1})\lambda_{2,n},a_1=\lambda_{2,1};b_n=(1+b_{n-1})\mu_{2,n},b_1=\mu_{2,1}
\eqno{(A-5.a)}
$$
and
$$
\lambda_2=(b_{N-1}-a_{N-1})\Delta t
\eqno{(A-5.b)}
$$
Similarly, we shall first concentrate on $a_n$. Before delving into the asymptotic behavior of  $a_n$ , we shall initially elucidate the interrelation between $\lambda_{2n}$ and $\lambda_{2,n+1}$  utilizing equation (16) 
$$
\lambda_{2,n+1}=\lambda_{2,n}\text{exp}(-\Delta B_{n} \Delta t\cdot i)=\lambda_{2,n}\text{exp}(O(\epsilon^{1+\alpha}\Delta t))
\eqno{(A-6)}
$$
Setting $\lambda_{2,0}=\lambda_{2,1}$,the approximation of $a_n$ is determined by
$$
a_n=\frac{\lambda_{2,n}}{1-\lambda_{2,n}} \left(1-\prod_{m=0}^{n-1}\lambda_{2,m}\right)+(n-1)O(\epsilon^{1+\alpha}\Delta t)
\eqno{(A-7)}
$$
We now employ mathematical induction to substantiate this conclusion.  For $n=1$, $a_1=\lambda_{2,1}=\frac{\lambda_{2,1}}{1-\lambda_{2,1}}(1-\lambda_{2,0})$ is trivial. Supposing that the aforementioned approximation is valid for $n$; hence, for the case of $n+1$, by recurrence relations in equation (A-5.a) we obtain
\begin{align*}
a_{n+1}=&(1+a_n)\lambda_{2,n+1}\\
=&\left(1+\frac{\lambda_{2,n}-\prod_{m=0}^{n} \lambda_{2,m}}{1-\lambda_{2,n}}+(n-1)O(\epsilon^{1+\alpha}\Delta t)\right)\lambda_{2,n+1}\\ 
=&\frac{\lambda_{2,n+1}}{1-\lambda_{2,n+1}}\left(1-\prod_{m=0}^{n}\lambda_{2,m}\right)
+\lambda_{2,n+1}\left(1-\prod_{m=0}^{n}\lambda_{2,m}\right) \left(\frac{1}{1-\lambda_{2,n}}-\frac{1}{1-\lambda_{2,n+1}}\right)\\&+(n-1)O(\epsilon^{1+\alpha}\Delta t)\lambda_{2,n+1}
\end{align*}
Since $|\lambda_{2,m}|=1$ holds for arbitrary $m$ , we have $\lambda_{2,n+1}(1-\prod_{m=0}^{n}\lambda_{2,m})=O(1)$  and $(n-1)O(\epsilon^{1+\alpha}\Delta t)\lambda_{2,n+1}=(n-1)O(\epsilon^{1+\alpha}\Delta t)$. The remaining term $(\frac{1}{1-\lambda_{2,n}}-\frac{1}{1-\lambda_{2,n+1}})$ can be reduced to
$$
\frac{1}{1-\lambda_{2n}}-\frac{1}{1-\lambda_{2,n+1}}=\left(\frac{1}{2\text{tan}\frac{\theta_n}{2}}-\frac{1}{2\text{tan}\frac{\theta_n+O(\epsilon^{1+\alpha}\Delta t)}{2}}\right)\cdot i=O(\epsilon^{1+\alpha}\Delta t)
\eqno{(A-8)}
$$
Integrating the preceding equations yields
$$
a_{n+1}=\frac{\lambda_{2,n+1}}{1-\lambda_{2,n+1}} \left(1-\prod_{m=0}^{n}\lambda_{2,m}\right)+n\cdot O(\epsilon^{1+\alpha}\Delta t)
\eqno{(A-9)}
$$
Thus the above proof is concluded. The estimation for $b_n$ yields an identical conclusion
$$
b_n=\frac{\mu_{2,n}}{1-\mu_{2,n}} \left(1-\prod_{m=0}^{n-1}\mu_{2,m}\right)+(n-1)O(\epsilon^{1+\alpha}\Delta t)
\eqno{(A-10)}
$$
To attain a structure akin to equation (19), it is imperative to fulfill the subsequent conditions
$$
(N-1)O(\epsilon^{1+\alpha}\Delta t)=O\left(\frac{T}{\Delta t}\cdot \epsilon^{1+\alpha}\Delta t\right)=O(\epsilon^{\alpha})
\eqno{(A-11)}
$$
and
$$
\prod_{m=0}^{N-1}\mu_{2,m}=\prod_{m=0}^{N-1}\lambda_{2,m}\text{exp}(-k\Delta t\cdot i)=(\prod_{m=0}^{N-1}\lambda_{2,m})\cdot \text{exp}(-kN\Delta t\cdot i)=\prod_{m=0}^{N-1}\lambda_{2,m}
\eqno{(A-12)}
$$
Combining equations (A-5.b),(A-7),(A-10),(A-11) and (A-12), the approximation of $\lambda_2$ is finally derived  
\begin{align*}
\lambda_2=&\left(\sum_{j=1}^{N-1} \prod_{n=1}^{j}  \mu_{2,N-n} - \sum_{j=1}^{N-1} \prod_{n=1}^{j}  \lambda_{2,N-n}\right)\Delta t\\
=&\left[\left(1-\prod_{m=0}^{N-2}\lambda_{2,m}\right)\left(\frac{\mu_{2,N-1}}{1-\mu_{2,N-1}}-\frac{\lambda_{2,N-1}}{1-\lambda_{2,N-1}}\right)+O(\epsilon^{\alpha})\right]\Delta t\\
=&\left[\left(1-\prod_{m=0}^{N-2}\lambda_{2,m}\right)\left(\frac{\lambda_{2,N-1}\text{exp}(-k\Delta t \cdot i)}{1-\lambda_{2,N-1}\text{exp}(-k\Delta t \cdot i)}-\frac{\lambda_{2,N-1}}{1-\lambda_{2,N-1}}\right)+O(\epsilon^{\alpha})\right]\Delta t
\end{align*}
which is identical to equation (19.c) by replacing $n$ with $N-n$ in $\lambda_{2,n}$ , as previously stated in (A-4).  

\section*{Acknowledgments}
~~~~J.W. thanks Zehua Qian and Youjun Hu for useful discussions. This work was supported by the National Natural Science Foundation of China under Grant No. 12205339.

\nocite{*}

\printbibliography
\end{document}